# On imaging capabilities of the Imaging and Medical beamline at the Australian Synchrotron

**T.E. Gureyev[a*], K.M. Pavlov[bcd], C.J. Hall[e], A. Maksimenko[e], D. Pelliccia[f], D.M. Paganin[c], A.W. Stevenson[e], A. Entezam[ae] and H.M. Quiney[a]**

[a] School of Physics, University of Melbourne, Parkville, Victoria, 3010, Australia
[b]School of Physical & Chemical Sciences, University of Canterbury, Christchurch, 8140, New Zealand
[c]School of Physics and Astronomy, Monash University, Clayton, Victoria, 3800, Australia
[d]School of Science and Technology, University of New England, Armidale, New South Wales, 2351, Australia
[e]Australian Synchrotron, Australian Nuclear Science and Technology Organisation, Clayton, Victoria, 3168, Australia
[f]Instruments & Data Tools Pty Ltd, Rowville, Victoria, 3178, Australia

[*]Correspondence email: timur.gureyev@unimelb.edu.au

**Synopsis** Several configurations for propagation-based and analyser-based phase-contrast imaging at the Imaging and Medical beamline of the Australian Synchrotron were examined both theoretically and experimentally.

**Abstract** Several propagation-based and analyser-based X-ray phase-contrast imaging experiments with over 100 m between the imaged objects and the detector were carried out at the Imaging and Medical beamline (IMBL) of the Australian Synchrotron. These experiments were aimed at characterization of possible phase-contrast imaging setups at IMBL. Effects of the beam divergence, polychromaticity and X-ray source size on the contrast and spatial resolution of images were evaluated quantitatively in different imaging scenarios. The results of this work will be used for improving the existing and developing new imaging capabilities at the beamline. Applications to X-ray phase-contrast tomography of biomedical samples, particularly for medical breast cancer imaging, are briefly discussed.

## 1. Introduction

X-ray phase-contrast imaging (PCI) exploits X-ray refraction to increase the quality of images of low-absorbing samples above that achievable with conventional absorption-based methods (Ando & Hosoya, 1972; Davis et al., 1995; Förster et al., 1980; Ingal & Beliaevskaya, 1995; Momose, 1995; Nugent et al., 1996; Snigirev et al., 1995; Somenkov et al., 1991; Wilkins et al., 1996). Over the years, multiple variants of X-ray PCI have been developed. Detailed descriptions and comparison of such methods can be found in several books and reviews (Endrizzi, 2018; Paganin, 2006; Wilkins et al., 2014). In the present paper, we investigated two variants of X-ray PCI, namely propagation-based imaging (PBI) (Nugent et al., 1996; Snigirev et al., 1995; Wilkins et al., 1996) and analyser-based imaging (ABI) (Ando & Hosoya, 1972; Bushuev et al., 1998; Chapman et al., 1997; Davis et al., 1995; Förster et al., 1980; Gureyev & Wilkins, 1997; Ingal & Beliaevskaya, 1995; Somenkov et al., 1991). The two techniques can be used at the Imaging and Medical beamline (IMBL) of the Australian Synchrotron (Stevenson et al., 2017) with the equipment readily available at the beamline. In the present paper, we study several variants of PBI and ABI that can be implemented in one of the three user hutches at IMBL. One of our main goals is a detailed evaluation of the beamline capabilities that are relevant for medical imaging in hutch 3B which is located at approximately 138 m from the X-ray source (Stevenson et al., 2017). This medical imaging mode has been used for development of three-dimensional breast cancer imaging at IMBL in the form of propagation-based phase-contrast computed tomography (PB-CT) (Arhatari et al., 2021; Gureyev et al., 2019). Research into medical lung imaging is also on-going in similar settings at IMBL (Costello et al., 2025). Other forms of medical X-ray imaging and radiation therapy are also being actively investigated at IMBL (Engels et al., 2025). The beamline can provide a coherent X-ray beam with width of up to 50 cm in hutch 3B, while the height of the beam is limited to approximately 4 cm (Hall, 2026; Stevenson et al., 2017). In order to enable projection imaging of a whole human female breast, a special component called a beam expander (BE) has been installed at IMBL. The BE can increase the height of the beam to approximately 8 cm in the targeted range of X-ray energies between approximately 32 and 38 keV. The BE utilizes one symmetric and one asymmetric Bragg reflection from Si crystals to expand the beam in the vertical direction. Details about the design and functionality of the BE are discussed in Section 2 below. While the main purpose of the BE is to provide a larger beam for imaging in hutch 3B, it can also be used as an analyser in ABI of samples imaged in hutches 1B and 2B which are located upstream of the BE. This form of imaging at IMBL is described in the present paper for the first time. However, since the BE is located at a distance of more than 90 m from hutch 3B where a detector was installed to enable the ABI experiments, the resultant imaging technique represented a combination of ABI and PBI, with both forms of phase contrast contributing to the images (Coan et al., 2005; Nesterets et al., 2005; Pavlov et al., 2004, 2005).

A prominent feature of the PBI and ABI experiments studied in the present paper is the use of a partially coherent divergent incident beam. Note that in this context one needs to distinguish between the “coherent” and “incoherent” divergence. The “coherent divergence” of the beam is determined by the angular spread of average propagation directions, with a unique propagation vector at each point of the quasi-spherical wavefront. The coherent divergence is determined primarily by the parameters of the electron beam in the storage ring of the synchrotron and by the superconducting wiggler, which is used as an insertion device generating the X-rays at IMBL (Stevenson et al., 2017). The coherent divergence is also modified by IMBL's double bent-crystal Laue monochromator (DBCLM) (Stevenson et al., 2017). The "incoherent divergence" is associated with the finite size of the (effectively incoherent) X-ray source (Pelliccia & Paganin, 2025) and can be characterized by a cone of different propagation directions at each point of the wavefront (Nugent, 2010). The X-ray source at IMBL has typical dimensions of approximately 800 μm horizontally and 40 μm vertically in terms of the full width at half-maximum (FWHM) (Stevenson et al., 2017). These dimensions of the X-ray source affect the spatial coherence of the X-ray beam. The nature of the effective temporal (chromatic) partial coherence is somewhat less straightforward. The DBCLM has an energy bandpass of approximately $\Delta\lambda / \lambda \sim 10^{-3}$ at wavelengths $\lambda$ relevant for medical imaging (Stevenson et al., 2017). On the other hand, the BE has a much narrower bandpass, of the order of $\Delta\lambda / \lambda \sim 10^{-6}$ (see details below). Therefore, the beam generated by the DBCLM is effectively polychromatic with respect to the wavelength bandpass of the BE. Combined with the coherent divergence of the beam, the interaction of the spatially and temporally partially coherent beam with the BE becomes non-trivial, as discussed in the next section. The overall picture of formation and propagation of the partially coherent divergent beam through various optical components at IMBL is complex, but its characterization is important for understanding of the beamline capabilities in ABI, PBI and PB-CT domains. Aspects of this picture are considered theoretically in the first parts of the present paper and later compared to the results of experiments. The main parameters of different possible experimental PCI setups and the characteristics of the BE are discussed in Section 2. Coherent image formation and geometrical magnification of images at IMBL are investigated in Section 3. In Section 4, we study partially coherent aspects of image formation, such as spatial resolution, while the basics of ABI and PBI contrast are outlined in Section 5 for completeness. Several key PCI experiments are described and analyzed in Section 6, followed by Conclusions in Section 7. Necessary details of the theory of X-ray diffraction by ideal crystals are included in the Appendix in order to make the paper largely self-contained.

## 2. Beam expander

The generic setting for PCI at IMBL is shown in Fig. 1. We carried out imaging experiments with samples in hutches 1B and 2B, at approximate distances of $R_{1,1B} = R_{1,1} + R_{1,2,1B} \cong 16$ m + 7 m =23 m and $R_{1,2B} = R_{1,1} + R_{1,2,2B} \cong 16$ m + 21 m =37 m from the nominal X-ray source position, respectively, where $R_{1,1} \cong 16$ m was the distance from the source to the DBCLM, while $R_{1,2,1B}$ and $R_{1,2,2B}$ were the distances from the DBCLM to the sample position in hutches 1B and 2B, respectively (see Fig. 1). In practice, the position of the effective (virtual) X-ray source is affected by the settings of the DBCLM, as well as the state of various slits along the beam path – see the discussion below. Other important geometric parameters of the imaging setups were the distances $R_{2,1,1B}$ and $R_{2,1,2B}$ between the imaged objects in hutches 1B and 2B, respectively, and the location of the BE. Finally, the distance between the BE and the fixed detector position in hutch 3B was denoted $R_{2,2}$ (Fig. 1). Most elements of the beamline were thoroughly characterized and described previously (Hall, 2026; Stevenson et al., 2017), with the exception of the BE. Therefore, our present work had a significant focus on the characterization of the BE and its performance in phase-contrast imaging. The BE is located at a distance of approximately 46 m downstream from the source and 98 m upstream from the fixed location where the detector was positioned in hutch 3B for our experiments. We also performed PBI and PB-CT experiments in the same setups, but with the BE removed from the beam. We mostly used the central X-ray energy $E$ = 35 keV in the experiments. The detector used was the IMBL Eiger2-3MW photon-counting detector with 75 μm ×75 μm pixels and a single-pixel point-spread function (PSF) (DECTRIS, 2026). Finally, we imaged some of the samples in hutch 3B at a short sample-to-detector distance for reference.

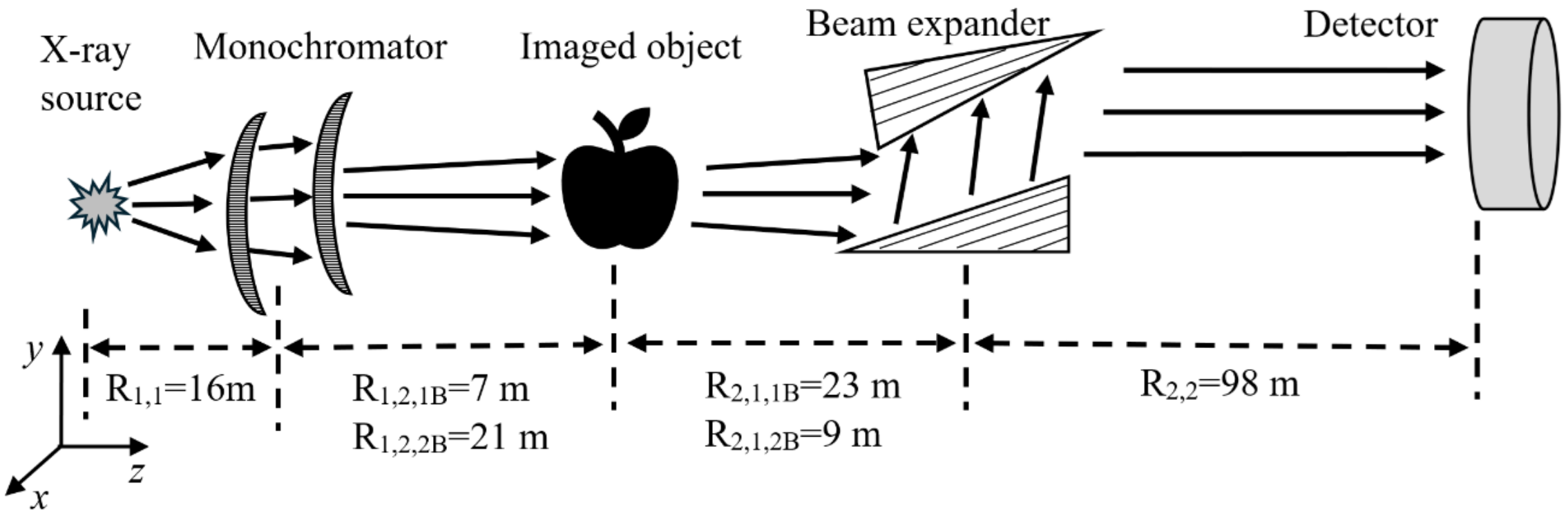

**Figure 1** Generic imaging setup of the experiments.

The BE at IMBL utilizes one symmetric and one asymmetric Si(333) crystal reflection in the vertical ($yz$) planes (Fig. 2). At the X-ray energy E = 35 keV, the Fourier coefficients of the crystal polarizability for Si(333) reflection are $\chi_0 = -0.78871 \times 10^{-6} + i \times 0.97551 \times 10^{-9}$, $|\chi_h| = 0.23073 \times 10^{-6}$ and the Bragg angle is $\theta_B$ = 9.7567 degrees (Pinsker, 1978; Stepanov, n.d.). In the second crystal, the (333) planes are cut at the angle $\phi$ = 8.503 degrees relative to the crystal surface (Fig. 2). This results in the following asymmetry factor (at 35 keV) (Fig. 3):

$$b = \frac{\gamma_0}{|\gamma_h|} = \frac{cos(\boldsymbol{k}_0,\boldsymbol{n})}{|\,cos(\boldsymbol{k}_h,\boldsymbol{n})|} = \frac{sin(\theta_B-\phi)}{|\,sin(\theta_B+\phi)|} \cong \frac{sin(9.757-8.503)}{sin(9.757+8.503)} \cong 0.06983. \qquad (1)$$

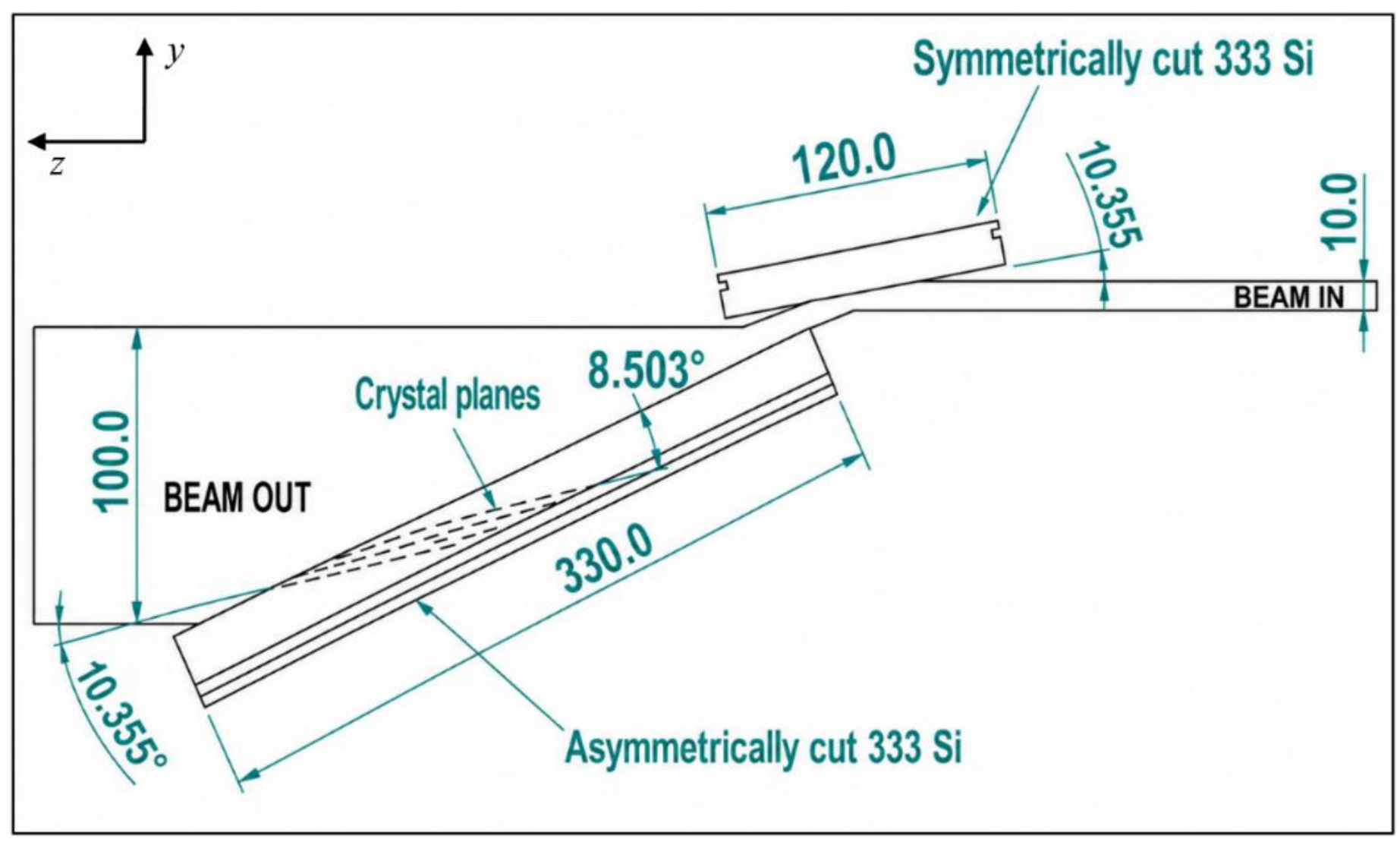

**Figure 2** Engineering drawing of the expected performance of the IMBL beam expander at E = 33 keV.

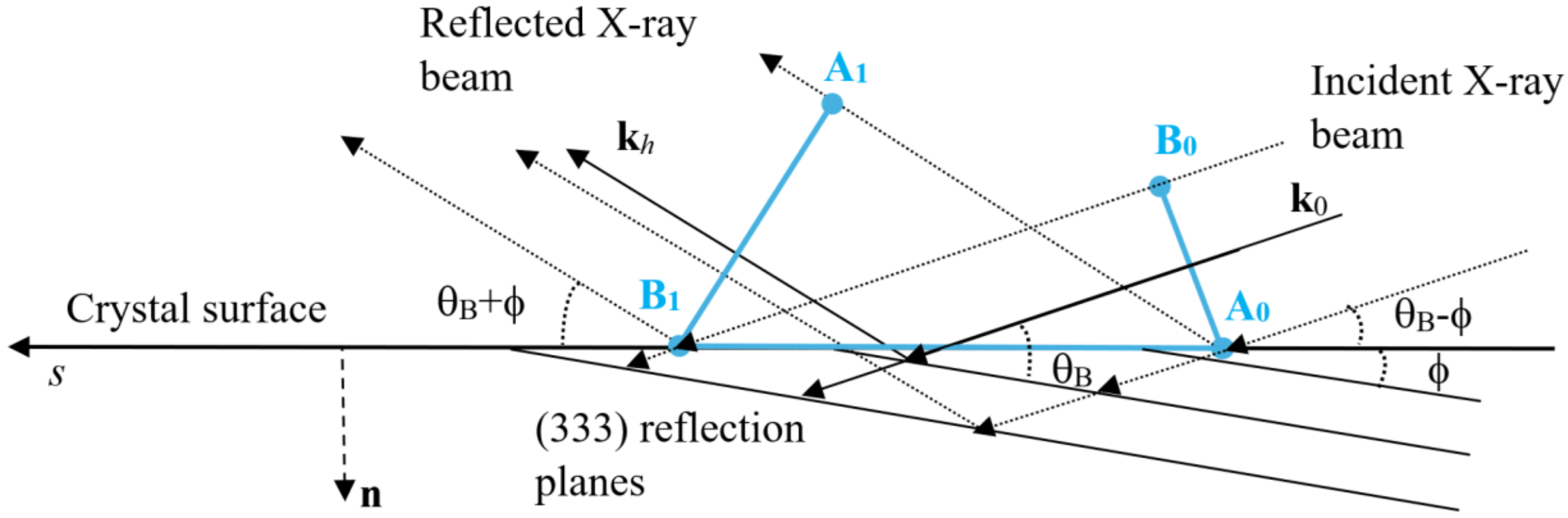

**Figure 3** Asymmetric Bragg reflection from a crystal (as for the second crystal of the BE): $\mathbf{k}_0$ – incident beam direction, $\mathbf{k}_h$ – reflected beam direction, **n** – normal to the crystal surface, $\theta_B$ – Bragg angle, $\phi$ - asymmetric cut angle, $(A_0,B_0)$ – transverse vertical cross-section of the incident beam, $(A_1,B_1)$ – transverse vertical cross-section of the reflected beam, $(A_0,B_1)$ – projection of the incident beam onto the crystal surface.

The linear increase of the vertical cross-section of the diffracted beam with respect to the incident beam is equal to $1/b$. Indeed, with reference to Fig. 3, we see that $|A_1B_1| = |A_0B_1|\sin(\theta_B+\phi) = [|A_0B_0|/\sin(\theta_B-\phi)]\sin(\theta_B+\phi) = |A_0B_0|/b$. This means that the magnification factor of the BE is equal to 1 / $b$ = 1.0 / 0.06983 ≅ 14.32 (at 35 keV). Correspondingly, after the BE, the intensity (X-ray flux) is multiplied by the factor of $b$, i.e. is reduced by approximately 14 times, according to energy (photon) conservation (ignoring the weak X-ray absorption in the BE). However, the actual reduction of the average beam intensity after reflection from the BE is much larger than $b$ due to the divergent and quasi-monochromatic nature of the X-ray beam produced by the DBCLM – see the discussion below. The coherent divergence of the beam is multiplied by the factor $b$ after the asymmetric reflection, according to the Liouville theorem (conservation of phase-space volume) (see also Appendix). This means that the beam becomes more collimated (less divergent) after the reflection from the BE.

The Darwin width of a rocking curve (RC) of a perfect crystal in the case of σ-polarization of the incident beam is equal to

$$W_{RC} = \frac{2|\chi_h|}{b^{1/2}\sin(2\theta_B)} \qquad (2)$$

(see Appendix for details). The first crystal of the BE utilizes a symmetric Si(333) reflection, which corresponds to eq.(2) with $b=1$. The RC width of the first crystal is equal to $W_1 \cong 0.285\ \text{arcsec} = 1.382$ µrad at E = 35 keV (Stepanov, n.d.). The RC width of the second crystal is equal to $W_2 \cong W_1/\sqrt{0.06983} = 1.0783\ \text{arcsec} = 5.228$ µrad. Provided that both crystals are aligned exactly at the Bragg angle, the width of the RC of the BE, $W_{BE}$, is approximately equal to the width of the narrower RC of the two, i.e. $W_{BE} \cong W_1 = 1.382$ µrad.

The coherent divergence and polychromaticity of the beam produced by IMBL's DBCLM dramatically increases the width of the effective RC of the BE. The experimentally measured width (FWHM) of the BE's RC at 35 keV was $W_{BE,poly}$ = 635 µrad (Fig. 4). This RC was obtained with the actual divergent polychromatic incident beam produced by the DBCLM, and with both crystals rotating synchronously in the $\theta$ - $2\theta$ geometry (Bowen & Tanner, 1998). We call this RC the polychromatic rocking curve of the BE for brevity.

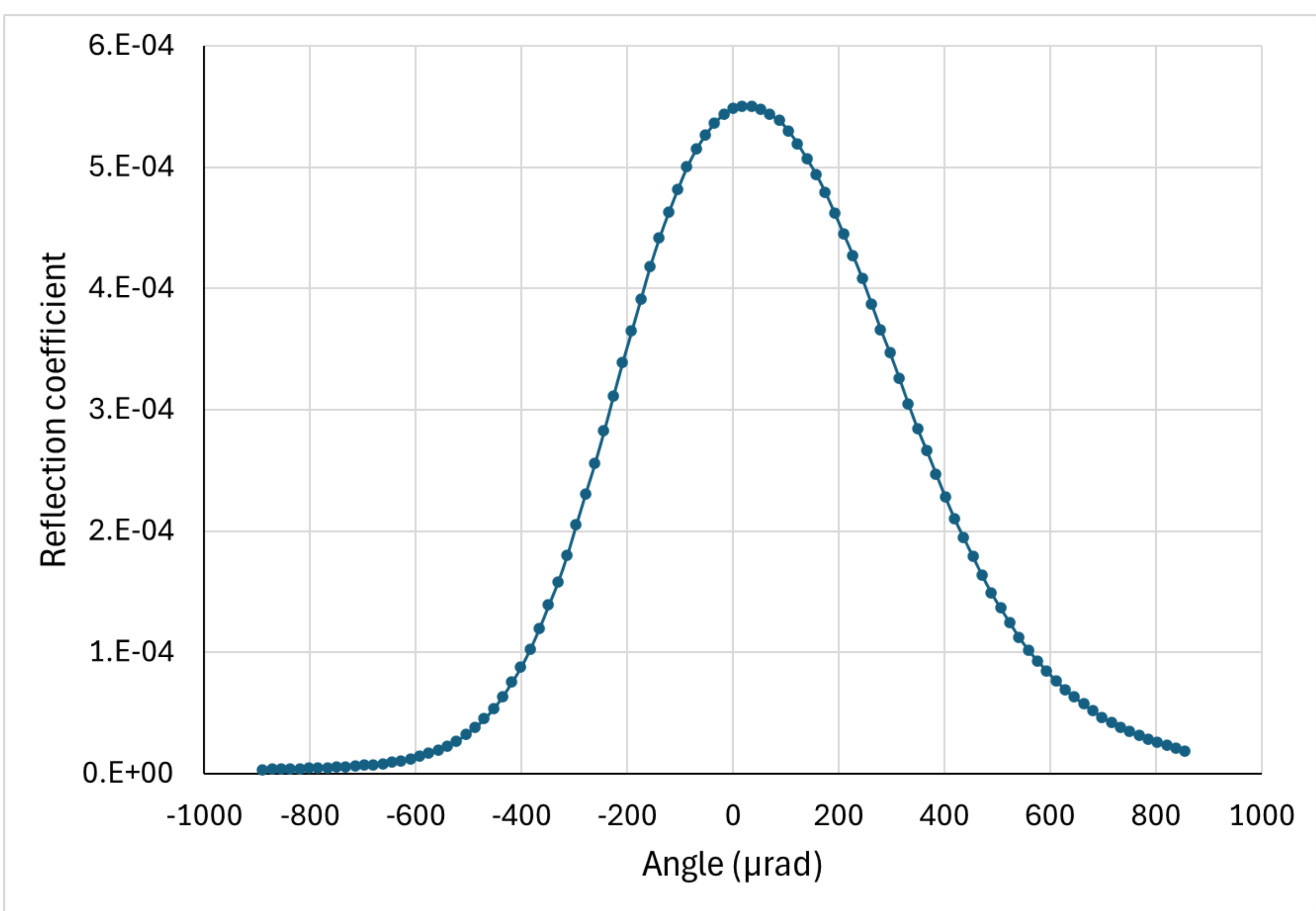

**Figure 4** Experimentally measured RC of the BE collected at IMBL with the beam produced by the DBCLM tuned at the central energy of $E = 35$ keV.

In the process of image formation in the case of a polychromatic incident beam, the RC of the BE is effectively convolved with the spectral-directional distribution of the beam incident on the BE, which is determined by the DBCLM (Stevenson et al., 2017). For narrow-band polychromatic X-ray beams, with $\Delta\lambda / \lambda << 1$, we can estimate the angular broadening, $\Delta\theta_\lambda$, of the RC due to polychromaticity using Bragg's law: $n\lambda = 2d_{ip} \sin\theta_B$ and $n(\lambda + \Delta\lambda) = 2d_{ip} \sin(\theta_B + \Delta\theta_\lambda) \cong 2d_{ip}(\sin\theta_B + \Delta\theta_\lambda \cos\theta_B)$, where $d_{ip}$ is the relevant interplanar distance of the crystal lattice. Subtracting the former equation from the latter, we get $n\Delta\lambda \cong 2d_{ip}\Delta\theta_\lambda \cos\theta_B$. Dividing the last equation by $n\lambda = 2d_{ip} \sin\theta_B$, we obtain an explicit expression,

$$\Delta\theta_\lambda \cong (\Delta\lambda/\lambda) \tan\theta_B, \tag{3}$$

for the angular broadening of the Bragg peak. The IMBL's DBCLM is known to produce an incident beam with spectral bandwidth of the order of $\Delta\lambda / \lambda \sim 10^{-3}$ at E = 35 keV (Stevenson et al., 2017).

The corresponding "polychromatic broadening" of the RC should then be equal to $\Delta\theta_\lambda = (\Delta\lambda/\lambda)\tan\theta_B \sim 10^{-3}\tan(9.757\text{ deg}) \cong 172$ μrad.

Considering the measured FWHM of the beam along the horizontal (*x*) and vertical (*y*) coordinate at different positions along the optical axis at IMBL (Hall, 2026) (see Table 1), it can be estimated that the coherent horizontal and vertical divergence angles of the beam at E = 35 keV were equal to $\Delta\theta_x \cong 5{,}410$ μrad and $\Delta\theta_y \cong 235$ μrad, respectively.

**Table 1** Linearly interpolated (over the beam energy) values of the measured width and height of the X-ray beam at different locations along the optical axis at IMBL at 35 keV.

| z (m) | Beam $FWHM_x$ (mm) | $\tan(\Delta\theta_x)$ | $\Delta\theta_x$ (μrad) | Beam $FWHM_y$ (mm) | $\tan(\Delta\theta_y)$ | $\Delta\theta_y$ (μrad) |
|---|---|---|---|---|---|---|
| 23 | 124.5 | 5.413E-03 | 5413 | 5.450 | 2.370E-04 | 237 |
| 38 | 205.5 | 5.408E-03 | 5408 | 8.925 | 2.349E-04 | 235 |
| 130 | 703.0 | 5.408E-03 | 5408 | 30.25 | 2.327E-04 | 233 |

The vertical coherent divergence range $\Delta\theta_y \cong 235$ μrad of the incident beam directly contributes to the width of the polychromatic RC of the BE. The horizontal divergence of the beam further increases the effective width of the RC of the BE, but the corresponding broadening is negligible in the present case: $(\Delta\theta_x/2)^2\tan\theta_B/2 \cong 0.629$ μrad (Beaumont & Hart, 1974; Pinsker, 1978). At each end (side) of the coherent divergence angular range, the RC is extended further due to the polychromatic broadening by the additional angular range $\Delta\theta_\lambda \cong 172$ μrad. This means that the expected value of the RC width is $W_{BE,poly} = \Delta\theta_y + 2\Delta\theta_\lambda \cong (235 + 2\times 172)$ μrad $\cong 579$ μrad. In order to obtain better agreement with the experimentally measured RC width of 635 μrad, we assume that the polychromatic broadening was closer to $\Delta\theta_\lambda = (635 - 235)/2 \cong 200$ μrad, which corresponds to the polychromaticity $\Delta\lambda/\lambda = \Delta\theta_\lambda/\tan\theta_B = 200$ μrad$/\tan(9.757\text{ deg}) \cong 1.163\times 10^{-3}$.

Let us consider the beam propagation in projection imaging at IMBL, during which the angular position of the BE is fixed. The expanding beam emerging from the DBCLM has a spectral bandwidth $\Delta\lambda$ around the central wavelength $\lambda$, with $\Delta\lambda/\lambda \cong 1.163\times 10^{-3}$ at E = 35 keV ($\lambda \cong 0.354$ Å). When this beam falls on the first crystal of the BE, only a small sub-range, $\lambda_y \pm \Delta\lambda_y$, of wavelengths present in the incident spectrum, is reflected from the crystal at each point *y* of its surface, where $\Delta\lambda_y/\lambda_y \le W_1/\tan\theta_B \cong 8.034\times 10^{-6}$. In principle, this range varies at different points *y*, because of the change of both the Bragg angle and the RC width as a function of the angle of incidence (which determines the central wavelength at a given point $y$). However, with the incident spectrum width $\Delta\lambda / \lambda \cong 1.163 \times 10^{-3}$, this variation is smaller than 1% of the range $\Delta\lambda_y / \lambda_y \cong 8.034 \times 10^{-6}$ (see Appendix) and hence can be neglected. On the other hand, the change of the central wavelength $\lambda_y$ at different points $y$ is relatively significant, as it corresponds to the full incident spectrum $\Delta\lambda / \lambda \cong 1.163 \times 10^{-3}$. Therefore, the reflected beam over the exit surface of the first BE crystal looks like a rainbow, with different "colors" (central wavelengths) reflected at different points of the crystal surface (Fig. 5). As the incident RC width of the second (asymmetric) crystal of the BE is broader than that of the first crystal, $W_2 > W_1$ (see eq.(2)), the "rainbow" beam is reflected by the second crystal essentially without a change in integral intensity or spatial color structure. However, as mentioned above, upon reflection from the second BE crystal, the beam is expanded by the factor $1 / b$ and, correspondingly, decreases in the average local intensity by the factor of $b$. The overall change of the average local intensity (fluence rate) of the beam produced by the DBCLM is therefore approximately equal to $b(\Delta\lambda_y / \lambda_y) / (\Delta\lambda / \lambda) \cong 6.983 \times 10^{-2} \times 8.034 \times 10^{-6} / 1.163 \times 10^{-3} = 4.824 \times 10^{-4}$ after the reflection from the BE. In other words, the average fluence rate of the beam decreases by a factor of approximately $1 / (4.824 \times 10^{-4}) \cong 2{,}073$ after reflection from the BE. The corresponding experimentally measured fluence rate reduction factor was found to be approximately 1,816 at E = 35 keV.

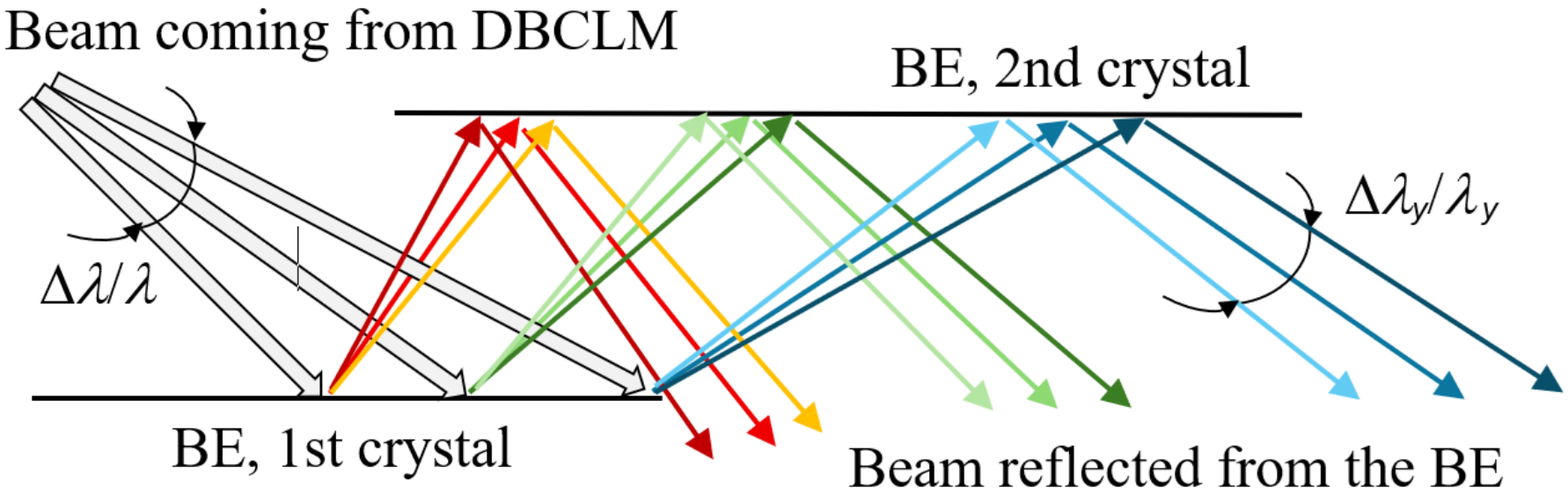

**Figure 5** Formation of the "rainbow" beam after reflection of a divergent quasi-monochromatic beam from the BE.

### *3.* Coherent imaging and geometrical magnification

When the BE is in use during an imaging experiment, geometrical magnification of objects imaged in hutches 1B or 2B in the setup of Fig. 1 is anisotropic, i.e. the magnification in the vertical ($yz$) planes is different from that in the horizontal (*xz*) planes (see Fig. 6). When the BE is not used, the geometrical magnification is isotropic.

In general, the geometrical magnifications of an imaged object at the detector plane can be calculated with reference to Fig. 6 as follows:

$X_3 = X_0(R_1 + R_2)/R_1 = X_0 M_x$,

$Y_1 = Y_0(R_1 + R_{2,1})/R_1$, $Y_2 = Y_1/b = Y_0(R_1 + R_{2,1})/(bR_1)$,

$Y_3 = Y_2 + R_{2,2} b Y_0/R_1 = Y_0(R_1 + R_{2,1} + b^2 R_{2,2})/(bR_1) = Y_0 M_y$,

where

$$M_x = (R_1 + R_2)/R_1 \text{ and } M_y = (R_1 + R_{2,1} + b^2 R_{2,2})/(bR_1). \quad (4)$$

The effective position of the virtual X-ray source is determined by the optical elements of the beamline, including the DBCLM and various slits. The distance $R_{1,2}$ between the DBCLM and the imaged object and the distance $R_{2,1}$ between the imaged object and the BE depend on the imaged object being located either in hutch 1B or hutch 2B (see Fig. 1).

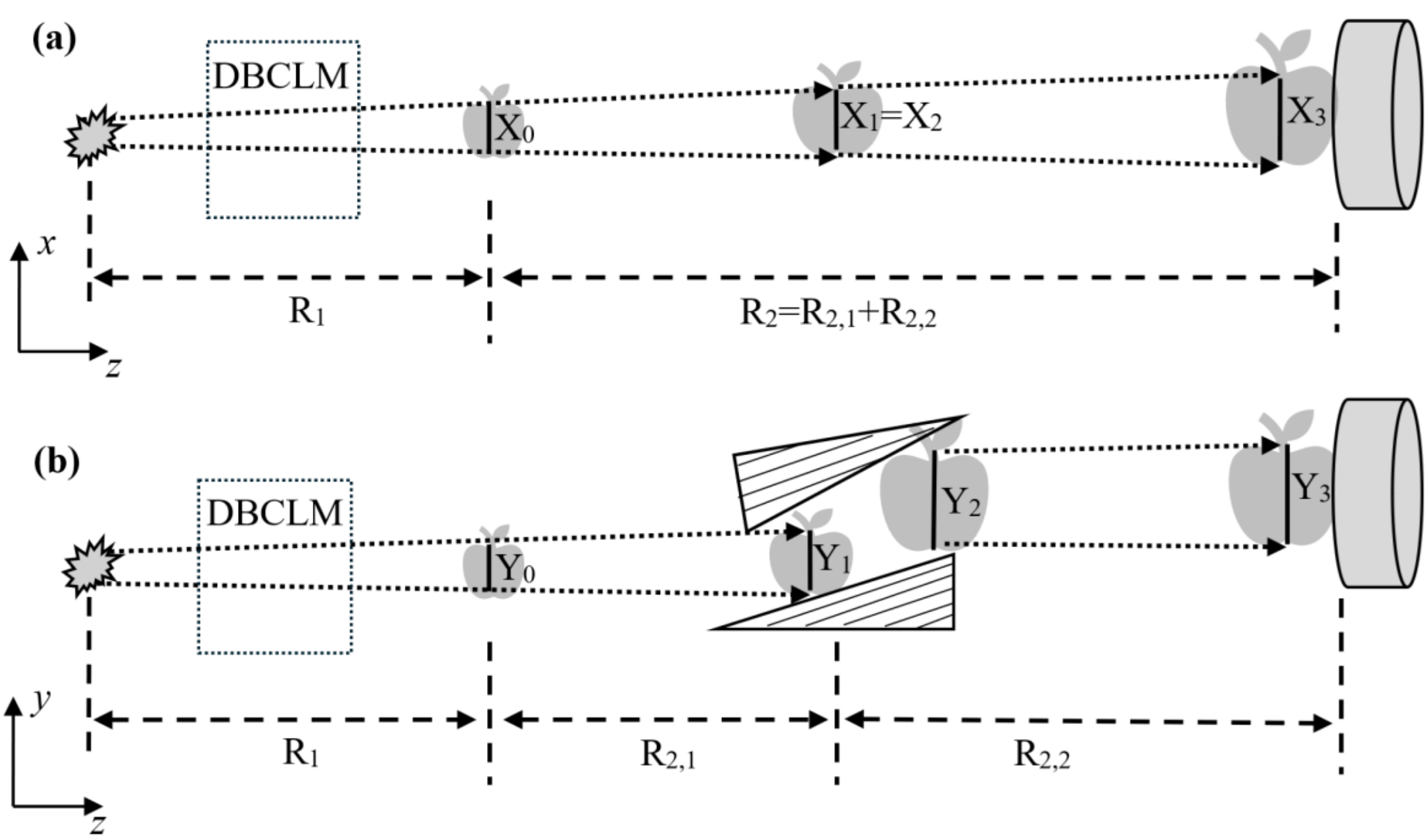

**Figure 6** Geometrical magnification setup at IMBL in the case of sample location in hutch 1B or 2B: (a) horizontal (*xz*) planes (no effects from the BE), and (b) vertical (*yz*) planes in the presence of the BE.

The different positions of the virtual source produced by the monochromator correspond to different curvature radii, $R_{curv}$, of the first (upstream) crystal in the monochromator (Suortti & Schulze, 1995). The calculations of such source positions were performed by applying an equation for a bent Laue crystal from Suortti et al. (1993), for two successive reflections in the DBCLM used at IMBL (Stevenson et al., 2017). The distance $z$ = 12.2 m corresponded to $R_{curv}$ = 6.94 m, which was calculated using the mechanical parameters of the bending in the case of a cold crystal. Heat load on DBCLM during its operation can produce a variation of the crystals' curvature (Wang et al., 2025). Taking this into account, a different radius of curvature, $R_{curv}$ = 10.4 m, was used for comparison with the experimental data. The latter radius of curvature corresponds to the position of the virtual source $z$ = 8.43 m, which fits well the experimentally measured data (see Table 2). When the beam size was significantly narrowed by slits, in order to improve the spatial resolution in the obtained images, it effectively produced a new virtual source at a position close to that of the DBCLM, $z$ = 16 m, which changed the magnification factors accordingly, as shown in Table 2. Note that these variations of the effective source position also affected the calculations of the spatial resolution below.

**Table 2** Theoretical and experimentally measured image magnification factors at IMBL at 35 keV. Bold font indicates the results corresponding to narrow slits.

| Magnification (M) | 3B, no BE | 3B, BE | 2B, no BE | 2B, BE | 1B, no BE | 1B, BE |
|---|---|---|---|---|---|---|
| Theor. M (source at z=8.43 m) | 1.046 | 1.001 | 4.745 | 19.071 | 9.305 | 37.396 |
| Theor. M (source at z=16 m) | **1.049** | **1.001** | **6.095** | **20.784** | **18.286** | **62.351** |
| Grid, period 363 µm | | | **6.3** | 19.3 | | 35.5 |
| Grid, period 313 µm | | | | 20.2 | | 35.9 |
| Cylinder, diam. 12.75 mm | | | 4.4 | | | |
| Metal wire, diam. 0.6 mm | | | | **20.0** | | |
| Steel aperture, diam. 27.9 mm | | | 3.9 | | | |
| Res. plate, period 0.1 mm | | | **6.2** | | | |
| Res. plate, period 0.4 mm | | | | **20.6** | | 39.4 |
| Res. plate, period 1.66 mm | | | | **20.0** | | |

### *4.* **Spatial resolution**

A rigorous description of image formation in ABI with polychromatic X-rays can be found, e.g., in Nesterets et al. (2005). Here we consider a simplified picture that allows one to estimate the spatial resolution in the imaging scheme shown in Fig. 1.

The ratio of geometrical magnifications of the X-ray source, $M_{s,x}$ and $M_{s,y}$, and the geometrical magnifications of the sample, $M_x$ and $M_y$ (see eq.(4)), can be expressed as:

$$M_{s,x} / M_x = (R_2 / R_1) / M_x = 1 - M_x^{-1} \text{ (without BE),} \tag{5a}$$

$$M_{s,y} / M_y = (R_{2,1} + b^2 R_{2,2}) / (bR_1 M_y) = 1 - (bM_y)^{-1} \text{ (with the BE).} \tag{5b}$$

Consequently, the effective source sizes in the object plane are

$$s_{o,x} = s_x M_{s,x} / M_x , \tag{6a}$$

$$s_{o,y} = s_y M_{s,y} / M_y , \tag{6b}$$

where $s_x$ and $s_y$ are, respectively, the horizontal and vertical size of the source in the source plane.

The effective (geometrically demagnified) detector pixel sizes in the object plane are equal to

$$d_{o,x} = d / M_x , \tag{7a}$$

$$d_{o,y} = d / M_y , \tag{7b}$$

where $d$ is the detector pixel size in the detector plane. Equations (7) correspond to the width of the detector PSF referred to the object plane in the case of a detector with a single-pixel PSF, such as e.g. the photon-counting Eiger2-3MW detector (DECTRIS, 2026) used in this experiment.

The effective X-ray source sizes (FWHM) at IMBL at 35 keV are known to be equal to $s_x \cong 800$ µm in the horizontal plane and $s_y \cong 40$ µm in the vertical plane (Hall, 2026; Stevenson et al., 2017). The Eiger2-3MW detector used in this experiment had a PSF with FWHM $d = 75$ µm. Substituting these values of the source size and the detector resolution into eqs.(6-7), we calculated the numerical values of the corresponding demagnified source sizes and detector resolutions for different imaging configurations (see Table 3). For completeness, we also included in Table 3 the contributions from the source size and detector resolution to the PSF in the image plane in the case of an object plane located in hutch 3B approximately 6 m upstream from the detector. The last imaging configuration is often used for biomedical imaging at IMBL (Arhatari et al., 2021; Gureyev et al., 2019).

The X-ray extinction length, $l_y$, in the BE (Pinsker, 1978) also contributes to image blurring in the vertical direction, limiting the vertical resolution in the object plane (see details in Appendix). Referred back to the object plane (see Fig. 6), the corresponding vertical image blurring width can be expressed as

$$l_{o,y} \cong \sqrt{1+b^3}\,\frac{1.22\lambda}{W_1}\,\frac{R_1}{R_1+R_{2,1}}. \tag{8}$$

Here we calculated the total extinction length in the BE as $l_y = [(l_y^{(1)}/b)^2 + (l_y^{(2)})^2]^{1/2}$, where $l_y^{(1)}$ and $l_y^{(2)} = b^{1/2} l_y^{(1)}$ are the extinction lengths of the first and second crystal of the BE, respectively, before projecting to the image plane and then referring back to the object plane. Note that the value $l_{o,y}$ in eq.(8) is reciprocal to the width of the (monochromatic) RC of the BE, rather than the width of the polychromatic RC.

The polychromaticity of the beam emerging from the DBCLM also contributes to image blurring when the BE is used. An incoherent divergence (dispersion) occurs when a polychromatic beam is reflected from an asymmetrically cut crystal (see Modregger et al., 2009, and the Appendix). The polychromaticity of X-rays incident on the first (symmetric) crystal of the BE is equal to $\Delta\lambda/\lambda \cong 1.163\times10^{-3}$. Because this reflection is symmetric, $b$ = 1, there is no additional incoherent beam divergence created at this reflection according to eq. (A8) of the Appendix. The polychromaticity of a parallel X-ray beam incident on the second (asymmetric) crystal of the BE is then limited by the bandpass of the first (symmetrical) Si(333) crystal of the BE, for which $\Delta\lambda/\lambda \le W_1/\tan\theta_B \cong 8.034\times10^{-6}$. When the beam reflected from the first crystal is then reflected from the second (asymmetric) crystal, it acquires an additional incoherent divergence equal to $\Delta\theta_{poly} = (1-b)W_1$ (see Modregger et al., 2009, and eq. (A8) in the Appendix). This leads to the image blurring height of $R_{2,2}\tan(\Delta\theta_{poly})$ in the image (detector) plane, which corresponds to the following blurring width in the object plane:

$$p_{o,y} = R_{2,2}\tan[(1-b)W_1]/M_y \tag{9}$$

for an object plane location upstream of the BE. Numerical values of the polychromatic blurring $p_{o,y}$ for different imaging configurations can be found in Table 3.

Taking into account the contributions from all the terms listed in eqs.(7-9), we can approximate the total FWHM of the PSF of the imaging system at the object plane as

$$h_{sys,x} = (s_{o,x}^2 + d_{o,x}^2)^{1/2}, \tag{10a}$$

$$h_{sys,y} = (s_{o,y}^2 + d_{o,y}^2 + l_{o,y}^2 + p_{o,y}^2)^{1/2} . \tag{10b}$$

Numerical values of $h_{sys,x}$ and $h_{sys,y}$ in different imaging configurations are given in Table 3.

**Table 3** Theoretical image blur contributions from the X-ray source size, BE and the detector resolution at IMBL at 35 keV. Bold font indicates the results corresponding to narrow slits.

| Object plane blur | 3B, no BE | 3B, BE | 2B, no BE | 2B, BE | 1B, no BE | 1B, BE |
|---|---|---|---|---|---|---|
| $s_{o,x}$ (z=8.43 m) (µm) | 35.4 | 35.4 | 631.4 | 631.4 | 714.0 | 714.0 |
| $s_{o,x}$ (z=16 m) (µm) | **1.9** | **1.9** | **33.4** | **33.4** | **37.8** | **37.8** |
| $s_{o,y}$ (z=8.43 m) (µm) | 1.8 | 0.4 | 31.6 | 10.0 | 35.7 | 24.7 |
| $s_{o,y}$ (z=16 m) (µm) | 1.9 | 0.6 | 33.4 | 12.4 | 37.8 | 30.8 |
| $d_{o,x}$ (z=8.43 m) (µm) | 71.7 | 71.7 | 15.8 | 15.8 | 8.1 | 8.1 |
| $d_{o,x}$ (z=16 m) (µm) | 71.5 | 71.5 | 12.3 | 12.3 | 4.1 | 4.1 |
| $d_{o,y}$ (z=8.43 m) (µm) | 71.7 | 74.9 | 15.8 | 3.9 | 8.1 | 2.0 |
| $d_{o,y}$ (z=16 m) (µm) | 71.5 | 74.9 | 12.3 | 3.6 | 4.1 | 1.2 |
| $l_{o,y}$ (z=8.43m) (µm) | 0.0 | 1.9 | 0.0 | 23.8 | 0.0 | 12.1 |
| $l_{o,y}$ (z=16 m) (µm) | 0.0 | 1.9 | 0.0 | 21.9 | 0.0 | 7.3 |
| $p_{o,y}$ (z=8.43 m) (µm) | 0.0 | 7.7 | 0.0 | 6.6 | 0.0 | 3.4 |
| $p_{o,y}$ (z=16 m) (µm) | 0.0 | 7.7 | 0.0 | 6.1 | 0.0 | 2.0 |
| $h_{sys,x}$ (z=8.43 m) (µm) | 79.9 | 79.9 | 631.6 | 631.6 | 714.1 | 714.1 |
| $h_{sys,y}$ (z=8.43 m) (µm) | 71.7 | 75.4 | 35.3 | 26.9 | 36.6 | 27.8 |
| $h_{sys,x}$ (z=16 m) (µm) | **71.5** | **71.5** | **35.6** | **35.6** | **38.0** | **38.0** |
| $h_{sys,y}$ (z=16 m) (µm) | 71.5 | 75.3 | 35.6 | 26.2 | 38.0 | 31.8 |

It can be seen from Table 3 that, when objects are imaged in hutch 3B, with magnification factors close to unity (see Table 2), the spatial resolution is approximately isotropic and is of the order of ~70-80 µm. This spatial resolution is largely determined by the detector resolution. On the other hand, the spatial resolution is of the order of ~600-750 µm in the horizontal direction and ~30-40 µm in the vertical direction, when objects are imaged in hutches 1B and 2B. These resolution values are dominated by the X-ray source size, because the corresponding magnification factors are significantly larger than unity (see Table 2). Since at IMBL the horizontal source size is 20 times larger than the vertical source size, the spatial resolution in hutches 1B and 2B is highly anisotropic and is very poor in the horizontal plane. As shown in some experimental examples below, it is possible to close down the slits to make the effective horizontal source size similar to the vertical source size in hutch 2B, however this results in a significant reduction of the horizontal field of view (beam width).

### *5.* Image contrast

The PBI contrast produced by a sharp feature in an imaged object can be defined as $C=(I_{\max}-I_{\min})/(I_{\max}+I_{\min})$, where $I_{\max}$ and $I_{\min}$ are the maximum intensity of the first bright Fresnel fringe and minimum intensity of the first dark Fresnel fringe, respectively, in the PBI image of the feature (Gureyev et al., 2008). It was shown by Gureyev et al. (2026) that, in the case of monochromatic X-rays with wavelength $\lambda$, the contrast generated by a homogeneous weakly-absorbing edge feature embedded in a bulk material can be expressed as

$$C(M,\lambda)\cong\frac{1-q(\lambda)}{1+q(\lambda)}-\frac{q(\lambda)\ln[q(\lambda)]}{(2\pi e)^{1/2}}\frac{\gamma(\lambda)}{N_{\mathrm{F}}}, \qquad (11)$$

where $q(\lambda)=\exp[-B(\lambda)]$, $B(\lambda)\equiv(4\pi/\lambda)\max_{x,y}\int\beta(x,y,z)dz<<1$ is the maximum X-ray absorption in the feature relative to the surrounding bulk material, $\gamma(\lambda)=\delta(x,z,\lambda)/\beta(x,z,\lambda)$ is the ratio of the real decrement to the imaginary part of the relative complex refractive index $n(x,z,\lambda)=1-\delta(x,z,\lambda)+i\beta(x,z,\lambda)$ of the homogeneous edge feature and $N_{\mathrm{F}}=\pi h_M^2/(R'\lambda)$ is the "minimal Fresnel number" associated with the characteristic width ("blurriness"), $h_M$, of the image of the edge referred back to the object plane. More specifically, $h_M^2=h_{\mathrm{sys}}^2+h_{\mathrm{obj}}^2$, where $h_{\mathrm{sys}}$ is the width of the PSF of the imaging system in the object plane, as expressed by eqs.(10), and $h_{\mathrm{obj}}$ is the "intrinsic width" of the edge feature itself. When the polychromaticity of the beam is of the order of $\Delta\lambda/\lambda\sim10^{-3}$, the effect of this polychromaticity on the contrast in eq.(11) is negligible. The first fraction on the right-hand side of eq.(11) corresponds to the conventional absorption contrast, while the second fraction corresponds to the PBI phase contrast. The PBI phase contrast term can be written as $C_{phase}\cong\Phi(\lambda)/N_{\mathrm{F}}$, where $\Phi(\lambda)\equiv(2\pi e)^{-1/2}\gamma(\lambda)B(\lambda)\exp[-B(\lambda)]$ is a dimensionless quantity depending only on the X-ray absorption and refraction characteristics of the object, while $1/N_{\mathrm{F}}=R'\lambda/(\pi h_M^2)$ depends only on the geometrical parameters of the imaging system and the X-ray wavelength. In typical cases of practical interest in PBI, it is usually desirable to maximize the PBI phase contrast by minimizing $N_{\mathrm{F}}$. This can be achieved by maximizing the effective propagation distance and/or minimizing the width of the PSF of the imaging system. Note, however, that eq.(11) is only valid under the condition $N_{\mathrm{F}} > \Phi(\lambda)$, with the phase contrast asymptoting to a constant value when $N_{\mathrm{F}}$ tends to zero (Gureyev et al., 2008, 2026). In the setups considered above, the effective propagation distance $R'$ can be different in the horizontal and vertical planes because of the effect of the BE:

$$R'_x = R_1 R_2 / (R_1 + R_2), \tag{12a}$$

$$R'_y = R_1 (R_{2,1} + b^{-2} R_{2,2}) / (R_1 + R_{2,1} + b^{-2} R_{2,2}). \tag{12b}$$

When $b^{-2} R_{2,2} >> R_1 + R_{2,1}$, as in the case of the experiments described in the present paper, we get $R'_y \cong R_1$. In order to simplify the theoretical analysis, we shall assume that the intrinsic width of the edge is much smaller than the spatial resolution of the imaging system (i.e. the edge is "sharp") and hence $h_M \cong h_{\mathrm{sys}}$. Using the expressions for the spatial resolution of the imaging system from eqs.(10), we obtain:

$$C_{phase,x} \cong \frac{\Phi(\lambda)\lambda R_1 R_2}{\pi (R_1 + R_2)(s_{o,x}^2 + d_{o,x}^2)}, \tag{13a}$$

$$C_{phase,y} \cong \frac{\Phi(\lambda)\lambda R_1}{\pi (s_{o,y}^2 + d_{o,y}^2 + l_{o,y}^2 + p_{o,y}^2)}. \tag{13b}$$

Equations (13) have a simple structure: the PBI contrast is directly proportional to the phase shift $\gamma(\lambda)B(\lambda)$ created by the edge, as well as to the effective propagation distance $R'$, and is inversely proportional to the square of the spatial resolution $h_{sys}$ of the imaging system. Note that for weakly absorbing features with $B(\lambda) << 1$ we have $\exp[-B(\lambda)] \cong 1 - B(\lambda)$ and hence $B(\lambda)\exp[-B(\lambda)] \cong B(\lambda) - B^2(\lambda) \cong B(\lambda)$, resulting in $\Phi(\lambda) \cong (2\pi e)^{-1/2} \gamma(\lambda) B(\lambda) = [2/(\pi e)]^{1/2} |\Delta\varphi(\lambda)|$, where $\Delta\varphi(\lambda)$ is the maximal phase shift generated by the edge feature. Nevertheless, the presence of the term $\exp[-B(\lambda)]$ in the expression for contrast in eq.(11) is essential in general for: (a) quantifying the contribution from absorption contrast, and (b) preventing the phase contrast term from formally "blowing up" in proportion to $B(\lambda)$ when the absorption becomes stronger. It may be also useful to note that the expression for phase contrast becomes very simple in the case of parallel-beam geometry and no BE: $C_x \cong [2/(\pi^3 e)]^{1/2} |\Delta\varphi(\lambda)| \lambda R_2 / d^2$ (Gureyev et al., 2026).

Table 4 gives examples of the theoretical values of PBI phase contrast $C_{phase} \cong \Phi(\lambda) / N_{\mathrm{F}}$, for a horizontally (along $x$) or vertically (along $y$) oriented 1 mm thick glandular tissue edge embedded in adipose tissue, in different imaging setups at IMBL at mean X-ray energy of 35 keV. In this example, we have the projected absorption $B(\lambda) \cong 6.9 \times 10^{-3}$, the "delta-to-beta" ratio $\gamma(\lambda) \cong 978$ (TS-Imaging, 2025) and the resultant phase shift $\Phi(\lambda) \cong 1.621$ rad. The values of the Fresnel number in different setups are included in Table 4. One can see that, depending on the setup, the PBI phase contrast values vary from fractions of a percent for the vertically oriented edges (horizontal contrast) in hutches 1B and 2B (both with and without the BE), to approximately 2% in all configurations with the sample in hutch 3B, and to values as high as 72% for a horizontally oriented edge (vertical contrast variation) in hutch 2B with the BE in use.

**Table 4** Theoretical phase contrast values for a 1 mm glandular tissue edge embedded in adipose tissue at different setups at IMBL at E = 35 keV. Bold font indicates the results corresponding to narrow slits.

| | 3B, no BE | 3B, BE | 2B, no BE | 2B, BE | 1B, no BE | 1B, BE |
|---|---|---|---|---|---|---|
| $R'_x$ (z=8.43 m) (m) | 5.7 | 5.7 | 22.5 | 22.5 | 13.0 | 13.0 |
| $R'_y$ (z=8.43 m) (m) | 5.7 | 6.0 | 22.5 | 28.5 | 13.0 | 14.6 |
| $R'_x$ (z=16 m) (m) | **5.7** | **5.7** | **17.6** | **17.6** | **6.6** | **6.6** |
| $R'_y$ (z=16 m) (m) | 5.7 | 6.0 | 17.6 | 21.0 | 6.6 | 7.0 |
| NF,x (z=8.43 m) | 98.8 | 98.8 | 1568.8 | 1568.8 | 3476.9 | 3476.9 |
| NF,y (z=8.43 m) | 79.5 | 84.0 | 4.9 | 2.3 | 9.1 | 4.7 |
| $NF_x$ (z=16 m) | **79.3** | **79.3** | **6.4** | **6.4** | **19.4** | **19.4** |
| $NF_y$ (z=16 m) | 79.3 | 84.0 | 6.4 | 2.9 | 19.4 | 12.8 |
| $C_{phase,x}$ (z=8.43 m) | 1.7E-02 | 1.7E-02 | 1.0E-03 | 1.0E-03 | 4.7E-04 | 4.7E-04 |
| $C_{phase,y}$ (z=8.43 m) | 2.1E-02 | 1.9E-02 | 3.3E-01 | 7.2E-01 | 1.8E-01 | 3.5E-01 |
| $C_{phase,x}$ (z=16 m) | **2.1E-02** | **2.1E-02** | **2.5E-01** | **2.5E-01** | **8.4E-02** | **8.4E-02** |
| $C_{phase,y}$ (z=16 m) | 2.1E-02 | 1.9E-02 | 2.5E-01 | 5.6E-01 | 8.4E-02 | 1.3E-01 |

ABI contrast can have a qualitatively different appearance depending on the ratio $\lambda^2 |\varphi''_{yy}(y)|/(\pi^2 W_{RC}^2)$ being much smaller or much larger than unity (Gureyev et al., 1997). Here and below, in the context of analysis of ABI imaging, we omit the coordinate *x* for brevity, since the ABI contrast generated by the IMBL BE appears only along the *y* coordinate; also, a dash denotes differentiation with respect to *y*. A sufficient validity condition for the "geometric optics" regime in ABI (Gureyev et al., 1997) is $|\varphi''_{yy}(y)| << \pi^2 W_{RC}^2 / \lambda^2$. Introducing the Takagi number, $N_T \equiv [\pi h W_{RC} / (2\lambda)]^2$ (Pavlov et al., 2004), where *h* is the spatial resolution in the object plane, a sufficient validity condition for the ABI geometric optics regime can be expressed in a simplified form:

$$|\varphi''_{yy}(y)| << 4N_T / h^2 . \tag{14}$$

In the case of coherent monochromatic light, we have $W_{RC} / \lambda \cong 1.22 / l_y$ (see eq.(A7) in the Appendix). Therefore, the Takagi number basically represents the squared ratio of the spatial resolution to the extinction length of the analyser crystal, $N_T \cong (2h / l_y)^2$. This is analogous to the Fresnel number being the squared ratio of the spatial resolution to the width of the first Fresnel zone, $\sqrt{R'\lambda}$. Note that the form of Takagi number used here corresponds to the limit, valid near the exact Bragg position, of a more general definition of the Takagi number originally introduced in (Pavlov et al., 2004) by analogy with the Fresnel number in the context of linearized imaging regimes in ABI and PBI.

In the case of monochromatic X-rays, if $\varphi(y)$ satisfies eq.(14), the ABI contrast can be expressed as (Gureyev et al., 1997)

$$C_{ABI} = \frac{|R[k^{-1}\varphi'(y_1,\lambda),\lambda] - R[k^{-1}\varphi'(y_2,\lambda),\lambda]|}{R[k^{-1}\varphi'(y_1,\lambda),\lambda] + R[k^{-1}\varphi'(y_2,\lambda),\lambda]}, \tag{15}$$

where $R(\Delta\theta;\lambda)$ is the rocking curve of the analyser crystal as a function of the deviation from the Bragg angle, $\Delta\theta = \lambda\eta$ and $\eta$ is the reciprocal space coordinate dual to *y* (see Appendix). In the monochromatic case, in order to achieve significant ABI contrast in eq.(15), it is necessary to have the the *y*-derivative of the phase divided by the wavenumber to differ, at two points $y_1$ and $y_2$, by a value comparable with or larger than the width of the RC of the analyser. In the polychromatic case, the situation is similar, but the role of the monochromatic RC, $R(\Delta\theta;\lambda)$, is now played by the polychromatic RC, $R_{poly}(\Delta\theta)$. The contrast can only be significant if the angular difference $\Delta\theta = | k^{-1}\varphi'(y_1,\lambda) - k^{-1}\varphi'(y_2,\lambda) |$ is comparable with or larger than the width of the polychromatic RC, which in the case of our experiment was equal to $W_{BE,poly} \cong 635$ µrad.

Let us consider a simple model sample in the form of a triangular wedge with a 90-degree top angle. After passing through the sample near the top of the wedge, the X-ray beam acquires a phase shift $\varphi(y,\lambda) = 2(2\pi/\lambda)\delta(\lambda)y\sin(45\deg)$, and the corresponding wavefront deviation angle is equal to $k^{-1}\varphi'(y,\lambda) = \sqrt{2}\,\delta(\lambda)$. For 35 keV X-rays and a wedge made of polycarbonate with $\delta(\lambda) \cong 2.158\times10^{-7}$, the deviation angle is equal to approximately 0.3 µrad, which is more than 1,000 times smaller than the width of the polychromatic RC of the BE. In order for a wedge-shaped sample to produce detectable ABI contrast in our considered setup, it is necessary either for the wedge angle to be extremely acute (less than ~0.1 degree in the previous example) or for the material of the wedge to shift the X-ray phase at 35 keV ~1000 times more strongly than polycarbonate. Note that while high-Z materials can have significantly larger values of $\delta(\lambda)$ compared to polycarbonate, they typically also have much higher X-ray absorption, which effectively decreases the observable phase contrast.

As another example, consider a model sample in the shape of a fibre (long solid cylinder) with radius $r$. If X-rays pass through the fibre near its edge, they acquire a phase shift $\varphi(y) = 2(2\pi/\lambda)\delta(\lambda)(r^2 - y^2)^{1/2}$. The corresponding deviation angle has a magnitude equal to $| k^{-1}\varphi'(y,\lambda) | = 2\delta(\lambda)y(r^2 - y^2)^{-1/2}$. In order for this magnitude to be comparable with $W_{BE,poly}$, the coordinate $y$ must satisfy the equation $2\delta(\lambda)y \cong W_{BE,poly}(r^2 - y^2)^{1/2}$ or $y \cong r[1 + 4\delta^2(\lambda)/W^2_{BE,poly}]^{-1/2}$. For example, at E = 35 keV and for a boron glass fibre, we have $\delta \cong 3.759\times10^{-7}$. As $2\delta(\lambda) \cong 7.518\times10^{-7} << W_{BE,poly}$, we can approximate the required value of $y$ as $y \cong r[1 - 4\delta^2(\lambda)/W^2_{BE,poly}] \cong 0.99999r$. For a fibre with radius of 1 mm, the points $y$ where the phase gradient reaches or exceeds the value required to produce detectable ABI contrast, are located within a distance of only ~0.7 nm from the edge of the fibre. This means that the corresponding phase contrast in the image can potentially be washed out by convolution with the PSF of the imaging system with a realistic width, such as e.g. 30-40 µm in the case of our experiment (see Table 3).

The preceding examples demonstrate that it may be difficult to obtain detectable ABI contrast in the imaging setups discussed above, with the IMBL BE in the role of the wavefront analyser, because the BE has a very broad (polychromatic) rocking curve. Nevertheless, experimental examples shown in the next section demonstrate that for certain types of objects it is still possible to register ABI contrast in the setup shown in Fig. 1. Additionally, in this type of imaging setups, with long distances between the object, the BE and the detector, the PBI and ABI contrasts can interact with each other, leading to a combined image contrast. Details of such combined ABI and PBI contrast formation have been previously described in several publications (see, e.g., Coan et al., 2005; Nesterets et al., 2005; Pavlov et al., 2004, 2005). As shown by Pavlov et al. (2004), for a weakly scattering object, the two contrasts are mostly additive, with the ABI resulting primarily in an area contrast and the PBI contrast more strongly expressed at edges and interfaces. The primary outcome of the interaction between the ABI and PBI contrasts is usually observed in the form of asymmetry of the PBI diffraction fringes at opposite edges of localized object features. Practical examples are considered in the next section.

### *6.* Experimental results

The main parameters of our experiments have been already described above - see Figs.1-6, and Tables 1-4, in particular. Here we discuss the results of imaging of several test objects in the imaging setups described above and compare these results with the theoretical estimates obtained in the previous sections. We collected multiple 2D projection images and CT scans of different samples positioned in hutches 1B, 2B and 3B, with the Eiger detector at a fixed position near the downstream wall of hutch 3B. Some of the images were collected with the BE and others without the BE in the beam. One goal of these experiments was to verify the geometrical image magnification factors obtained in different setups. The outcomes of the measurements of the magnification factors are shown in Table 2, with the experimental measurements generally being in agreement with the theoretical values obtained using the known positions of the X-ray source, the samples and the detector, as well as the magnification due to the asymmetric reflection of the beam from the BE. It turned out that the largest uncertainty in these measurements was associated with the effective position of the X-ray source, i.e. the effective point of origin of the coherent divergence of the beam. This position depended on the location of IMBL's insertion device, as well as on the effects of the DBCLM and various slits on the beam (Hall, 2026). According to the measurements, in a typical setting, the effective source position was located at approximately $z = 8.43$ m. However, when the slits were closed down sufficiently to make the horizontal spatial resolution comparable with the vertical one, the effective source position shifted to approximately $z = 16$ m, as confirmed by the measurements (see Table 2 and examples below).

#### *6.1.* Steel meshes

We imaged two different metal wire meshes with periods of 313 μm and 363 μm in hutches 1B, 2B and 3B at IMBL. Some of the images of the wire meshes collected in different setups are shown in Fig. 7. The "contact" images, collected at a sample-to-detector distance of 20 cm in hutch 3B are shown in Fig. 7(a-b). These images allowed us to verify the periods of the meshes using the known pixel size of the detector. The images collected in hutches 2B and 1B with the sample-to-detector distances of 107 m and 121 m, respectively, are shown in Figs. 7(c-e). These images were obtained with the BE in the beam. Since the BE expands the beam only in the vertical direction, the vertical and horizontal magnifications in these images were quite different (see Table 2 for details). As the effective source sizes in the horizontal and vertical directions were also very different, the spatial resolution was highly anisotropic in these images. As can be seen in Table 3, the $x$-resolution in hutches 1B and 2B was of the order of ~600-700 µm in the studied image settings, which was comparable to two periods of the meshes. Therefore, the mesh structure was completely washed out in the horizontal direction in the images shown in Fig. 7(c-e), while the wire structure was clearly visible in the vertical direction, where the spatial resolution was of the order of ~30 µm. Only when we closed down the slits to the extent that the effective horizontal source size became comparable with the vertical one, did the mesh structure become clearly visible in both the horizontal and the vertical directions (Fig. 7(f)). Measuring the mesh periods in the images shown in Fig. 7, we calculated the apparent geometrical magnifications in the considered setups, with the results shown in Table 2. We also imaged a metal plate with line patterns with different number of lines per mm, and then measured the apparent geometrical magnifications in those images as well. The results were consistent with those obtained from the images of the wire meshes (see Table 2).

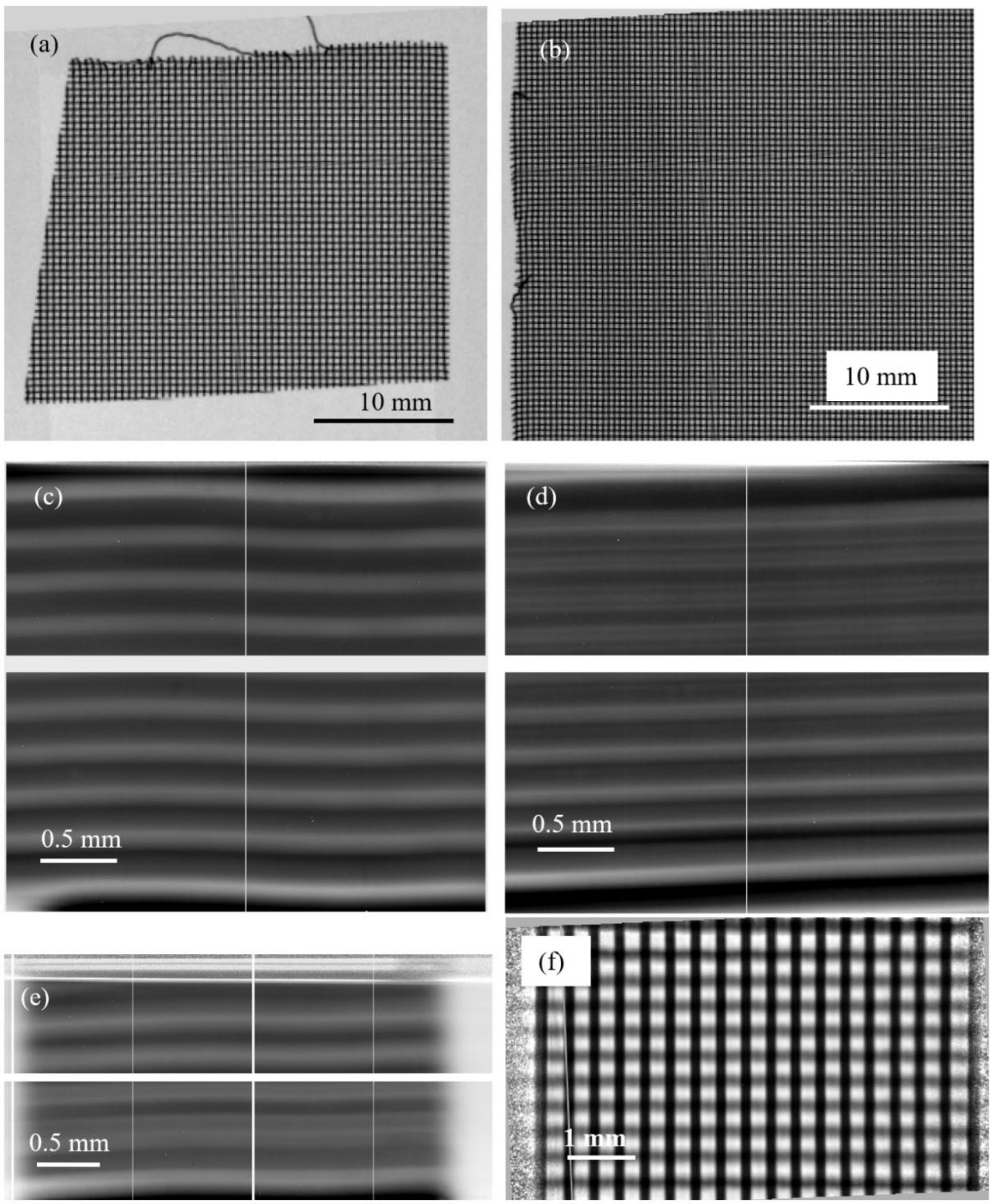

**Figure 7** Images of metal wire meshes with periods of 363 μm, (a), and 313 μm, (b), located in hutch 3B at 20 cm from the Eiger detector. The meshes with 363 μm period, (c), and 313 μm, (d), imaged in hutch 2B, with the BE and with the Eiger detector in hutch 3B. (e) The mesh with 363 μm period in hutch 1B, imaged with the BE and with the Eiger detector in hutch 3B. (f) The mesh with 363 μm period imaged in hutch 2B with the Eiger detector in hutch 3B, without the BE and with the horizontal source size minimized by the slits.

### *6.2.* Glass capillaries

Another test sample imaged in this experiment consisted of a solid nickel-copper alloy wire with a diameter of 0.6 mm and two hollow boron glass capillaries with outer diameters of 0.3 mm and 0.1 mm and wall thickness of 10 µm, mounted inside a circular aperture with a diameter of 27.9 mm in an aluminium plate. As the wire and capillaries were mounted horizontally, this sample could be used for producing and assessing the image contrast in the vertical ($y$) direction. When this sample was imaged in hutch 3B at 20 cm from the Eiger detector (Figs.8(a-b)), the two capillaries produced only weak absorption contrast with a maximum of around 1%, while the wire exhibited the maximum absorption of approximately 74%. When the same sample was imaged in hutch 2B, with the detector at a distance of 107 m from the sample, and with the BE in the beam, phase contrast became clearly visible in the images (Fig. 8(c-d)). As can be seen in Figs.8(c-d), the wire predominantly produced PBI contrast in the form of Fresnel fringes near the edges. The images of both capillaries in Figs.8(c-d) clearly show a combination of the PBI and ABI contrast. The ABI contrast, produced due to the reflection of the transmitted beam from the BE, appeared in the form of an intensity gradient (slope) along the $y$ coordinate, in accordance with the phase gradient of the transmitted beam. A remarkable feature of the images of the capillaries in Figs.8(c-d) was that the 10 µm glass wall was clearly resolved, which was particularly obvious in the image of the smaller capillary in Figs.8(c-d). This

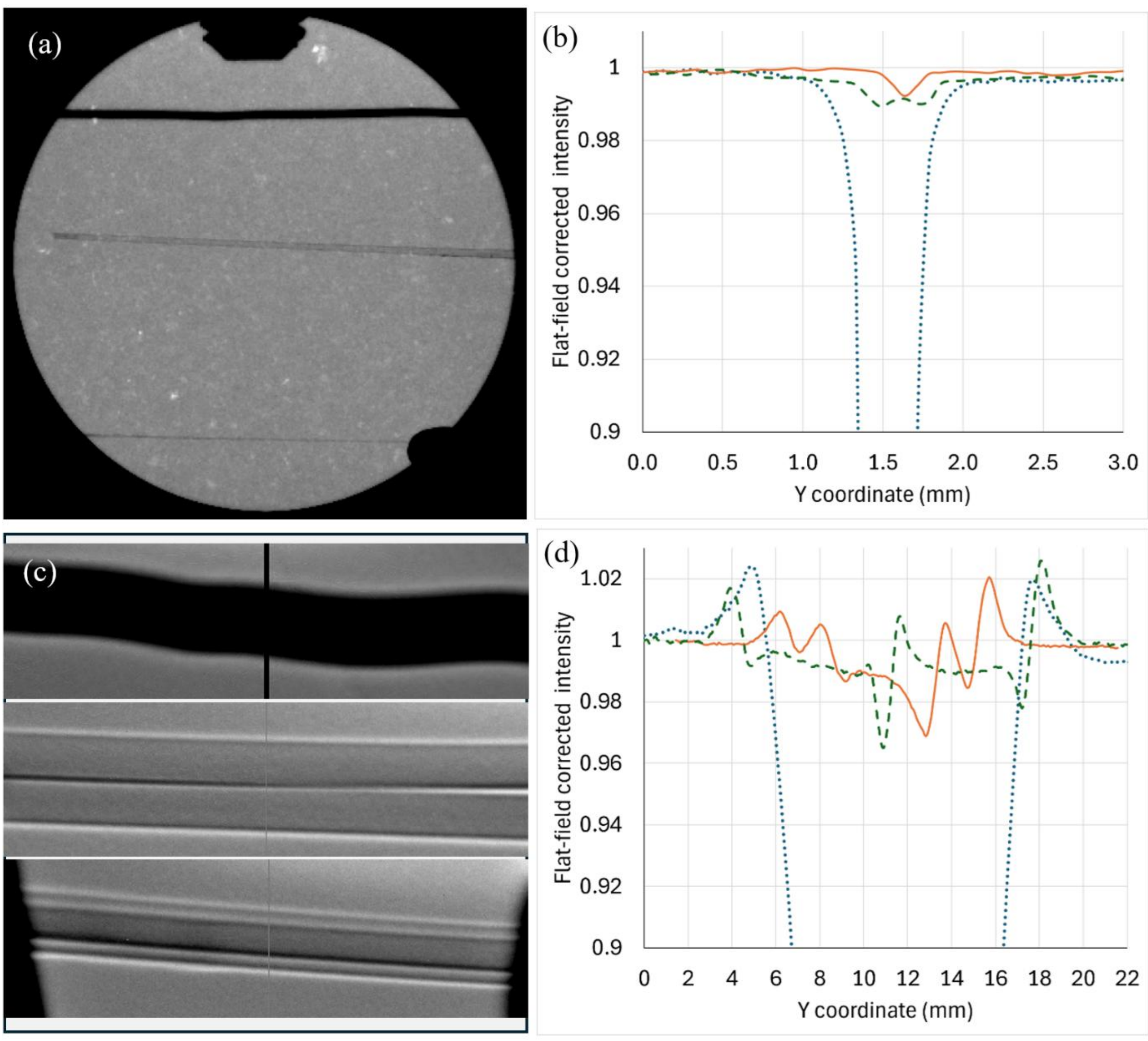

**Figure 8** (a) Contact image of a wire (top) and two boron glass capillaries with outer diameters of 0.3 mm (middle) and 0.1 mm (bottom), mounted inside a circular aperture with diameter 27.9 mm in an aluminium plate. (b) Vertical line profiles through the wire and two capillaries shown in (a): wire - dotted blue line, 0.1 mm capillary – solid orange line, 0.3 mm capillary – dashed green line. (c) Fragments of images of the same wire and two capillaries as in (a), but imaged at 107 m sample-to-detector distance with the BE. (d) Vertical line profiles through the wire and two capillaries shown in (c) using the same line styles as in (b).

"super-resolution" was achieved due to the large magnification factor in this image and the very strong phase contrast which was not washed out completely by the relatively broad PSF of the imaging system with the width of approximately 30 μm (see Table 3). The effective geometrical magnification in the horizontal direction was measured using the image of the aperture in the aluminium plate. The measured magnification factor $M_x \cong 4.2$ was in approximate agreement with the theoretical horizontal magnification ("no BE") in hutch 2B in the case of the source at $z$ = 8.43 m (see Table 2). The effective geometrical magnification in the vertical direction was measured by comparing the width of the wire in Figs.8(a-b) and in Figs.8(c-d). The measured value $M_y \cong 20.0$ was consistent with the theoretical vertical magnification in hutch 2B with the BE and the source at $z$ = 8.43 m (see Table 2). Due to the additional divergence of the transmitted X-ray wavefront generated by the capillaries, their effective magnification in the vertical direction observed in Fig. 8(c-d) was equal to approximately 43 for the larger capillary and 85 for the smaller capillary. The magnitude of the combined ABI and PBI contrast in Figs.8(c-d) was up to 4% for the smaller capillary and up to 8% for the larger capillary. The observed improvement of the visibility of the details of the structure of the capillaries in Figs.8(c-d) in comparison with the "contact" image in Figs.8(a-b) was quite significant, as a result of the combined effects of image magnification, improvement in the effective spatial resolution and phase contrast.

### *6.3.* Breast tissue samples in paraffin

We also collected several PB-CT scans at E = 35 keV of samples mounted on a CT rotation stage in hutch 2B at 107 m from the Eiger detector in hutch 3B, without the BE in the beam. For these scans, we closed down the slits to the extent that made the spatial resolution in the horizontal direction comparable to that in the vertical direction (see Fig. 7(f)). This led to a substantial reduction in the horizontal field of view, but, as the following examples demonstrate, it was still possible to collect images with a horizontal field of view of up to ~35 mm. Six of these PB-CT scans were performed using paraffin block samples that had a parallelepiped shape with approximately 40 mm × 25 mm in the horizontal cross-sections and 10 mm in height. The paraffin blocks contained samples of breast tissue with cancerous lesions. These samples were imaged in accordance with the Human Ethics Certificate of Approval from Monash University Human Research Ethics Committee. Such paraffin-embedded tissue samples are often used in standard histopathological workflows (Pati et al., 2024). The images reconstructed from two PB-CT scans of the paraffin block samples are shown in Fig. 9. The geometrical magnification in these images was approximately 6.3 in both the horizontal and vertical directions (see Table 2). The theoretical spatial resolution was approximately 35.6 µm in both directions (see Table 3). Measurements of spatial resolution in the reconstructed PB-CT slices consistently produced values close to 3 detector pixels, i.e. 3 × 75 µm / 6.3 ≅ 35.7 µm. As one can see in Fig. 9, the reconstructed PB-CT images demonstrated a high contrast of breast tissues, which was routinely of the order of 15% for spiculated soft tissue features, reaching as high as 25% for some of the features shown in Fig. 9(a). It is also noteworthy that microcalcifications (such as the ones shown in Fig. 9(a) and (c)), with the size of the order of 200 μm or smaller, had the contrast of the order of 35% in these images. Clear visualization of microcalcifications is important because of their high association with cancerous lesions in breast cancer (Aminzadeh et al., 2022). Also important is the ready availability of quantitative 3D information in these images, which can potentially make them useful as an adjunct tool in the histopathological workflow (Baran et al., 2018; Pati et al., 2024).

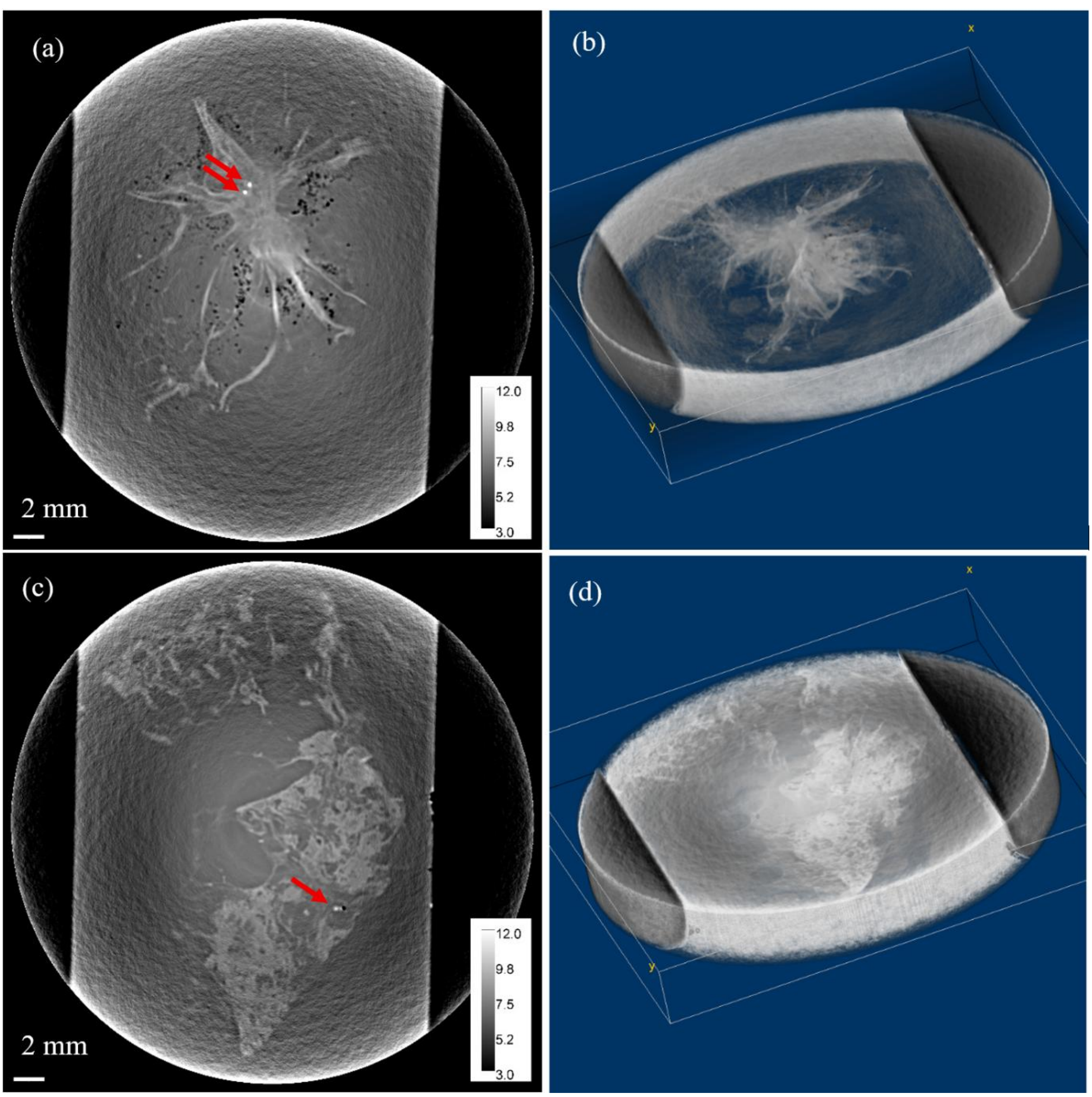

**Figure 9** PB-CT images of two breast tissue samples embedded in paraffin. Both samples were imaged in hutch 2B with the Eiger detector in hutch 3B (sample-to-detector distance 107 m). Red arrows indicate microcalcifications. (a) Reconstructed distribution of $\beta \times 10^{11}$ in a coronal slice of sample 1. (b) 3D rendering of PB-CT reconstruction of sample 1. (c) Reconstructed distribution of $\beta \times 10^{11}$ in a coronal slice of sample 2. (d) 3D rendering of PB-CT reconstruction of sample 2.

### 7. Conclusions

We have investigated several different scenarios for phase-contrast imaging at IMBL. Our theoretical calculations of the achievable contrast and spatial resolution have confirmed that, in the case of large biomedical samples, comparable in size with major human organs, the best conditions are achieved with both the sample and the detector located in hutch 3B. By allowing a large distance (over 130 m) from the X-ray source, it is possible to achieve a large illumination area (transverse section of the beam), with up to 50 cm horizontally and 4 cm vertically in the X-ray energy range relevant to medical imaging (roughly, between 30 and 80 keV). In the range of approximately 30-40 keV it is also possible to use the beam expander to enlarge the vertical size of the beam up to approximately 8 cm in hutch 3B, enabling imaging of a whole human breast in a single parallel-beam CT scan with a circular scanning trajectory over 180 degrees. The last setting is going to be used for breast cancer imaging of live patients at IMBL in the near future (Arhatari et al., 2021; Gureyev et al., 2019). With the long source-to-sample distance of ~140 m, which is significantly longer than the maximum available sample-to-detector distance of ~7 m in hutch 3B, the source is effectively demagnified in the images and the spatial coherence of the beam becomes high. Furthermore, due to the source demagnification, the very high anisotropy (40 μm height vs 800 μm width) of the IMBL X-ray source does not affect the spatial resolution in any significant way in this imaging mode, and the resolution remains approximately isotropic if the detector has an isotropic PSF. In that scenario, the spatial resolution in the images is determined primarily by the detector, as long as the spatial resolution of the detector is larger than approximately 70 μm, which is suitable for medical imaging modalities, such as breast and lung imaging.

While the PBI contrast in the above imaging setup in hutch 3B can be adequate (typically, of the order of a few percent for soft tissue features of interest) for medical imaging applications, we have shown that the contrast can be made significantly higher in a different PBI mode, where the samples are placed in hutches 1B or 2B at a distance of over 100 m from the detector in hutch 3B. We also showed that in this mode it is possible to improve the overall spatial resolution approximately two-fold compared to imaging with both the sample and the detector in hutch 3B. More specifically, we showed that in the case of PBI at E = 35 keV of human soft tissue samples in hutch 2B, using the Eiger detector with the spatial resolution of 75 μm in hutch 3B, it is possible to achieve phase contrast of the order of 30%. Also, in this case, when using a detector with 75 μm spatial resolution, the spatial resolution in the object plane can be approximately 35 μm in the vertical direction, i.e. twice finer than in the case of PBI imaging in hutch 3B. The main problem with PBI imaging in hutches 1B and 2B is related to the horizontal X-ray source size, which is equal to ~800 μm. When the geometrical magnification is high and the overall spatial resolution of the imaging system is dominated by the source size, the resolution becomes highly anisotropic at IMBL, being quite poor in the horizontal direction. It is possible to improve the horizontal resolution and reduce the anisotropy by closing down the slits upstream of the object. We have demonstrated that it is possible to equalize the effective horizontal resolution with the vertical resolution in this way. Unfortunately, closing of the slits in such a manner significantly reduces the horizontal beam size, i.e. the field of view. We have demonstrated that it is still possible in practice to perform PBI imaging of samples with the diameter of up to ~3 cm in hutch 2B, with isotropic spatial resolution of the order of 35 μm.

Regarding the options for ABI imaging at IMBL, using the BE as the analyser crystal, we established that in general this modality is unlikely to be very useful in most biomedical imaging use cases. The main issue in this case is represented by the very broad rocking curve of the BE in the context of divergent "polychromatic" beam (with monochromaticity of the order of $10^{-3}$) generated by the IMBL's monochromator (DBCLM). Due to the combination of the coherent divergence of the beam and its polychromaticity, the BE is able to reflect incident X-rays with different wavelengths at different incident angles. This increases the width of the RC and effectively washes out ABI contrast that can be generated by typical biological samples. Note however that the same broad RC of the BE makes it effective in its primary function, i.e. in expanding the incident beam coming from the DBCLM in the vertical direction. The broad RC allows the BE to accept incident X-rays with different energies from the total range of $\Delta\lambda/\lambda \sim 10^{-3}$ and different incident directions within the total coherent divergence of approximately 235 μrad, to produce a collimated and vertically expanded beam. We noted that the expanded beam has a "rainbow" structure, i.e. the central wavelength (colour) changes within the overall range of $\Delta\lambda/\lambda \sim 10^{-3}$ as a function of the vertical coordinate, while at any given height the local monochromaticity (temporal coherence) is of the order of $\Delta\lambda/\lambda \sim 10^{-5}$. Fortunately, this peculiar chromatic structure of the expanded beam is largely irrelevant for biomedical imaging in hutch 3B, as the polychromaticity of the order of $\Delta\lambda/\lambda \sim 10^{-3}$ is usually negligible in biomedical PBI image formation. However, a fact of significant importance in this case is the reduction of the average intensity (fluence rate) of the beam by a factor of approximately 2,000, compared to the unexpanded beam. We explained that this reduction factor is the result of combination of two factors: (1) the expansion of the beam by a factor of approximately 14 due to the asymmetric reflection from the second crystal of the BE, and (2) the additional monochromatization of the incident beam by a factor of approximately 144 by the BE. As a consequence, while the intensity of the beam from DBCLM in hutch 2B can deliver the dose rate to soft tissues of ~0.4 Gy / s, the expanded beam delivers the maximum dose rate of the order of 0.2 mGy / s in hutch 3B. The latter dose rate is just high enough for practical breast cancer imaging, allowing a CT scan with a ~3 mGy mean glandular dose to be completed within ~15 s, i.e. within a time compatible with the realistic maximum breast-hold duration for a typical patient. Future upgrades to IMBL may include measures for increasing the power of the insertion device (brilliance of the X-ray source) and improvements to the throughput efficiency of beamline elements such as the monochromator, the BE and various filters, in order to increase the dose rate delivered by the expanded beam in biomedical imaging in hutch 3B.

**Acknowledgements** The authors are grateful to A/Prof Marcus Kitchen of Monash University for the supply of several samples used in the experiments in this work and for helpful discussions.

**Conflicts of interest** The authors declare no conflicts of interest in relation to this publication.

**Data availability** The simulated and most of the experimental data used in this paper will be made available upon reasonable request. The breast tissue scans data cannot be made available due to patient privacy regulations.

**Funding information**

National Health and Medical Research Council (grant No. APP2011204).

***Appendix A.* Bragg reflection of an X-ray wave from a perfect crystal**

In the case of a thick centrosymmetric crystal and σ-polarization, the reflection of a plane X-ray wave from the crystal in the Bragg geometry can be described by the following expression (Afanas'ev & Kohn, 1971):

$$U_{out}(s;A) = i(b\chi_h / \chi_{\bar{h}})^{1/2} \int_0^s \exp(iAs')[J_1(Bs')/s']U_{in}(s-s')ds' , \qquad \text{(A1)}$$

where $U_{in}(s)$ and $U_{out}(s;A)$ are the complex amplitudes of the incident and reflected beams, respectively, $s$ is the coordinate running along the surface of the crystal in the plane of diffraction, $J_1(s)$ is the Bessel function of the first kind and of the first order,

$$A = k\chi_0\gamma_0 \frac{(1+b^{-1})}{2\sin(2\theta_B)} - k\alpha \frac{\gamma_0}{2\sin(2\theta_B)} = k\gamma_0(\Delta\theta + \theta_0) ,\ \alpha = -2\Delta\theta\sin(2\theta_B) ,\ b = \gamma_0 / |\gamma_h| ,$$

$$\theta_0 = \frac{\chi_0(1+b^{-1})}{2\sin(2\theta_B)} \text{ and } B = \frac{k|\chi_h|\sqrt{\gamma_0|\gamma_h|}}{\sin(2\theta_B)} = \frac{k\gamma_0|\chi_h|}{b^{1/2}\sin(2\theta_B)} .$$

In order to establish the link between the coordinates in the plane orthogonal to the direction of the beam incident on the crystal and those along the surface of the crystal (see Fig. 3), we note that $As = 2\pi(\zeta + \zeta_0)s$ , where $2\pi\zeta = k\gamma_0\Delta\theta$ , $\zeta = \Delta\theta\gamma_0 / \lambda$ and $\zeta_0 = \theta_0\gamma_0 / \lambda$ . If $\eta$ is reciprocal to $y$ in the transverse planes of the incident beam, then on the crystal surface we have $y = s\gamma_0$ , $\eta = \zeta / \gamma_0 = \Delta\theta / \lambda$ , $\eta_0 \equiv \zeta_0 / \gamma_0 = \theta_0 / \lambda$ and $A = 2\pi\gamma_0(\eta + \eta_0)$ . If the height of the incident beam (along the $y$ coordinate) is $Y_{in}$, it creates a support (area of non-zero disturbance) with length equal to $S_{in} = Y_{in} / \gamma_0$ for the complex amplitude on the crystal surface $U_{in}(s)$ . It can be seen from eq.(A1) that the support $S_{out}$ of $U_{out}(s)$ generally coincides with the support of $U_{in}(s)$ , with a relatively small addition of an extra length of the order of $s_{ext}$ ("extinction length" – see below) at one end, i.e. $S_{out} \cong Y_{in} / \gamma_0$ . For the coordinate $y'$ in the transverse planes of the reflected beam, we have $y' = s|\gamma_h| = y / b$ . Therefore, the width of the reflected beam is $Y_{out} = S_{out}|\gamma_h| \cong Y_{in}|\gamma_h| / \gamma_0 = Y_{in} / b$ . This means that after an asymmetric reflection with $b < 1$ (as in the case considered in the main text of this paper), the width of the beam increases by the factor $1/b$ . Applying the same line of reasoning to the Fourier transforms of $U_{in}(s)$ and $U_{out}(s)$, and the projections of their respective supports onto the reciprocal coordinates $\eta$ and $\eta'$ , we obtain that the width of the Fourier transform of $U_{in}(s)$ , i.e. the coherent divergence of the beam around the $y$ direction, decreases by the factor $b$ upon the reflection.

The complex amplitude reflection coefficient, $r(A)$, corresponds to the limit case $s = +\infty$ in eq.(A1), $U_{out}(+\infty; A) = r(A)U_{in}$. In the case of a non-absorbing crystal, it has the following form (Bushuev et al., 1998; Pinsker, 1978):

$$r(A) = b^{1/2}\left[\eta_B + \mathrm{sgn}(1+\eta_B)\sqrt{\eta_B^2 - 1}\right]^{-1}, \tag{A2}$$

where $\eta_B = \dfrac{\alpha_B b^{1/2}}{2|\chi_h|} = \dfrac{Ab^{1/2}}{|\chi_h|}\dfrac{\sin(2\theta_B)}{k\gamma_0} = \dfrac{A}{B}$ and

$\alpha_B = -\alpha + \chi_0(1+b^{-1}) = 2\Delta\theta\sin(2\theta_B) + \chi_0(1+b^{-1}) = 2A\sin(2\theta_B)/(k\gamma_0)$. We can also write eq. (A2) in terms of the coordinate $\Delta\theta = \lambda\eta$ by noting that

$\eta_B = \dfrac{A}{B} = \dfrac{2(\Delta\theta + \theta_0)}{W_{RC}} = \dfrac{2\lambda(\eta + \eta_0)}{W_{RC}} = \dfrac{2\lambda(\zeta + \zeta_0)}{\gamma_0 W_{RC}}$, where

$$W_{RC} = \frac{2|\chi_h|}{b^{1/2}\sin(2\theta_B)}, \tag{A3}$$

see, e.g., Pinsker (1978). Substituting this into eq.(A2), we obtain

$$r(\Delta\theta) = \frac{b^{1/2}W_{RC}}{2}\left[\Delta\theta + \theta_0 + \mathrm{sgn}(1+\eta_B)\sqrt{(\Delta\theta+\theta_0)^2 - W_{RC}^2/4}\right]^{-1}. \tag{A4}$$

Note that $W_{RC}$ can be associated with the width of the RC measured in $\Delta\theta$. Indeed, the interval $|\Delta\theta + \theta_0| \le W_{RC}/2$ corresponds to the interval $-1 \le \eta_B \le 1$. The RC is, by definition, equal to $R(\eta_B) \equiv |r(\eta_B)|^2$ and hence can be expressed from eq. (A2) as follows (Pinsker, 1978):

$$R(\eta_B) = b \times \begin{cases} 1, \text{ when } -1 \le \eta_B \le 1 \; (-W_{RC}/2 \le \Delta\theta + \theta_0 \le W_{RC}/2), \\ \left(|\eta_B| + \sqrt{\eta_B^2 - 1}\right)^{-2}, \text{when} |\eta_B| > 1 \; (|\Delta\theta + \theta_0| > W_{RC}/2). \end{cases} \tag{A5}$$

Therefore, $W_{RC}$ corresponds to the angular width of the "Darwin table", i.e. the flat central region of the RC of a non-absorbing perfect crystal.

The "extinction length" along the coordinate $s$ on the surface of the crystal can be defined from eq.(A1) by the condition $Bs_{ext} = s_{1,0}$, where $s_{1,0} \cong 3.832$ is the location of the first zero of the function $J_1(s)$ (Nesterets et al., 2004). Then

$$s_{ext} = \frac{3.832}{B} = \frac{3.832\, b^{1/2}\sin(2\theta_B)}{k\gamma_0|\chi_h|} \cong \frac{7.663}{k\gamma_0 W_{RC}}. \tag{A6}$$

When projected onto the $y$ axis, $y = s\gamma_0$, this gives the following expression for the extinction length projected onto $y$:

$$l_y = s_{ext}\gamma_0 \cong \frac{7.663}{kW_{RC}} \cong \frac{1.220\lambda}{W_{RC}}. \qquad \text{(A7)}$$

Let us now analyze how some of the physical quantities considered above change within a narrow range of wavelengths, $\Delta\lambda / \lambda \sim 10^{-3}$.

Table A1. Si(333) reflection parameters for 34.9, 35.0 and 35.1 keV.

| $E$ (keV) | $\lambda$ (Å) | $\theta_B$ (deg) | $\lvert\chi_0\rvert$ | $\lvert\chi_h\rvert$ |
|---|---|---|---|---|
| 34.9 | 0.3553 | 9.7850 | $0.7932\times10^{-6}$ | $0.2321\times10^{-6}$ |
| 35 | 0.3543 | 9.7567 | $0.7887\times10^{-6}$ | $0.2307\times10^{-6}$ |
| 35.1 | 0.3532 | 9.7287 | $0.7842\times10^{-6}$ | $0.2294\times10^{-6}$ |

For the Si(333) reflection, $b = 1$ and $b \cong 0.0698$, using values from Table A1, we obtain the following values for the combined extinction length of the first and second crystals of the BE: $L_y(34.9\,\text{keV}) \cong 32.458\,\mu\text{m}$, $L_y(35\,\text{keV}) \cong 32.435\,\mu\text{m}$ and $L_y(35.1\,\text{keV}) \cong 32.415\,\mu\text{m}$. We see that the extinction lengths differ by less than 0.1 μm within this energy range. We are also interested in the change of the RC as a function of $\lambda$, for $\Delta\lambda / \lambda \sim 10^{-3}$, around the central wavelength $\lambda = 0.3543\times10^{-4}\,\mu\text{m}$ (corresponding to $E$ = 35 keV). For $b$ = 1, we get: $W_{RC}(34.9\,\text{keV}) \cong 1.386\,\mu\text{rad}$, $W_{RC}(35.0\,\text{keV}) \cong 1.382\,\mu\text{rad}$ and $W_{RC}(35.1\,\text{keV}) \cong 1.377\,\mu\text{rad}$. Therefore, the widths of the (monochromatic) RCs differ less than by 0.01 μrad within this energy bandwidth. The variation of the position of the center of the RC, $\theta_0 = 0.5\chi_0(1+b^{-1}) / \sin(2\theta_B)$, within the interval between $E$ = 34.9 keV and $E$ = 35.1 keV, in the case $b$ = 1, is described by the difference between $\lvert\theta_0(34.9\,\text{keV})\rvert \cong 2.366\,\mu\text{rad}$ and $\lvert\theta_0(35.1\,\text{keV})\rvert \cong 2.352\,\mu\text{rad}$, $\lvert\theta_0(34.9\,\text{keV}) - \theta_0(35.1\,\text{keV})\rvert \cong 0.014\,\mu\text{rad}$, which is a negligibly small number compared to $W_{RC}$.

When a parallel polychromatic beam is Bragg-reflected from an asymmetrically cut crystal, it acquires an additional divergence ("dispersion") (Modregger et al., 2009):

$$\Delta\theta_{ad} = (1-b)(\Delta\lambda / \lambda)\tan\theta_B. \qquad \text{(A8)}$$

This happens because the conservation of momentum equation, $\mathbf{k}_h = \mathbf{k}_0 + \mathbf{h}$, when projected onto the crystal surface, implies (see Fig. 3):

$$k\cos(\theta_B - \phi) - (2\pi / d)\sin\phi = k\cos(\theta_B + \phi), \qquad \text{(A9)}$$

and, for $\lambda_1 = \lambda - \Delta\lambda$, we have $k_1 = 2\pi / (\lambda - \Delta\lambda) \cong k(1 + \Delta\lambda / \lambda)$ and hence

$$k(1 + \Delta\lambda / \lambda)\cos(\theta_B - \phi) - (2\pi / d)\sin\phi = k(1 + \Delta\lambda / \lambda)\cos(\theta_B + \phi - \Delta\theta_{ad}). \qquad \text{(A10)}$$

Let us show that $\Delta\theta_{ad}$ is generally not equal to zero. Substituting $\cos(\theta_B + \phi - \Delta\theta_{ad}) \cong \cos(\theta_B + \phi) + \sin(\theta_B + \phi)\Delta\theta_{ad}$ into eq.(A10) and subtracting eq.(A9) from it, we obtain $(\Delta\lambda / \lambda)\cos(\theta_B - \phi) \cong (\Delta\lambda / \lambda)\cos(\theta_B + \phi) + \sin(\theta_B + \phi)\Delta\theta_{ad}$, or $\Delta\theta_{ad} \cong (\Delta\lambda / \lambda)[\cos(\theta_B - \phi) - \cos(\theta_B + \phi)] / \sin(\theta_B + \phi)$. Since $\cos(\theta_B - \phi) - \cos(\theta_B + \phi) = 2\sin\theta_B \sin\phi$, we arrive at $\Delta\theta_{ad} \cong 2(\Delta\lambda / \lambda)\sin\theta_B \sin\phi / \sin(\theta_B + \phi)$. Furthermore,

$$\begin{aligned}
&2\sin\theta_B \sin\phi / \sin(\theta_B + \phi) = 2\tan\theta_B[\cos\theta_B \sin\phi / \sin(\theta_B + \phi)] = \\
&\tan\theta_B[(\sin(\theta_B + \phi) - \sin(\theta_B - \phi)) / \sin(\theta_B + \phi)] = \\
&\tan\theta_B[1 - \sin(\theta_B - \phi) / \sin(\theta_B + \phi)] = (1 - b)\tan\theta_B,
\end{aligned}$$

which proves eq. (A8).